\documentclass[aps,pre,amsmath,amsfonts,amssymb,superscriptaddress]{revtex4}
\usepackage{graphicx}

\let\a=\alpha \let\b=\beta  \let\d=\delta
\let\e=\varepsilon  \let\h=\eta 
\let\l=\lambda \let\m=\mu   
\let\s=\sigma \let\t=\tau  
   \let\G=\Gamma
\let\D=\Delta   
   
 \let\r=\rho  \let\io=\infty

\def\NN{{\cal N}} 
 \def\JJ{{\cal J}} 
\def\GG{{\cal G}}

\def\to{\rightarrow} \def\la{\left\langle} \def\ra{\right\rangle}

\newcommand{\wh}{\widehat} 
\newcommand{\Tr}{\text{Tr}}

\def\dd{{\rm d}}
\def\un{{\underline{n}}}

\def\u0{{\underline{0}}}

\def\bn{{\boldsymbol{n}}}
\def\ubn{{\underline{\boldsymbol{n}}}}
\def\uy{{\underline{y}}}
\def\by{{\boldsymbol{y}}}
\def\uby{{\underline{\boldsymbol{y}}}}

\def\Ns{{N_{\rm s}}}
\def\Nt{{{\cal N}_{\rm traj}}}
\def\Ntry{{{\cal N}_{\rm tries}}}

\def\hW{\widehat{W}}

\def\Tr{{\rm Tr}}

\def\ba{{\boldsymbol{a}}}
\def\bac{{\boldsymbol{\bar{a}}}}
\def\ac{{\bar{a}}}

\def\bPhi{{\boldsymbol{\Phi}}}
\def\bPsi{{\boldsymbol{\Psi}}}

\def\I{{\rm 1\hspace{-0.90ex}1}}

\def\Z{{\cal Z}}
\def\di{{\partial i}}
\def\dimj{{\partial i \setminus j}}
\def\ij{{\langle i,j\rangle }}
\def\ik{{\langle i,k\rangle }}
\newcommand{\beq}{\begin{equation}} 
\newcommand{\eeq}{\end{equation}}
\newcommand{\bea}{\begin{eqnarray}} 
\newcommand{\eea}{\end{eqnarray}}

\begin{document}

\title{Exact solution of the Bose-Hubbard model on the Bethe lattice}

\author{Guilhem Semerjian}
\affiliation{LPTENS, Unit\'e Mixte de Recherche (UMR 8549) du CNRS et
  de l'ENS, associ\'ee \`a l'UPMC Univ Paris 06, 24 Rue Lhomond, 75231
  Paris Cedex 05, France.}

\author{Marco Tarzia}
\affiliation{Laboratoire de Physique Th\'eorique de la Mati\`ere Condens\'ee, 
Universit\'e Pierre et Marie Curie-Paris 6, UMR CNRS 7600, 
4 place Jussieu, 75252 Paris Cedex 05, France.}

\author{Francesco Zamponi}
\affiliation{LPTENS, Unit\'e Mixte de Recherche (UMR 8549) du CNRS et
  de l'ENS, associ\'ee \`a l'UPMC Univ Paris 06, 24 Rue Lhomond, 75231
  Paris Cedex 05, France.}

\begin{abstract}

The exact solution of a quantum Bethe lattice model in the thermodynamic limit
amounts to solve a functional self-consistent equation. In this paper we 
obtain this equation for the Bose-Hubbard model on the Bethe lattice, under
two equivalent forms. The first one, based on a coherent state path integral,
leads in the large connectivity limit to the mean field treatment of 
Fisher et al. [Phys. Rev. B {\bf 40}, 546 (1989)] at the leading order, 
and to the bosonic Dynamical Mean Field Theory as a first correction,
as recently derived by Byczuk and Vollhardt 
[Phys. Rev. B {\bf 77}, 235106 (2008)].
We obtain an alternative form of the equation using the occupation number
representation, which can be easily solved with an arbitrary numerical 
precision, for any finite connectivity.
We thus compute the transition line between the superfluid  
and Mott insulator phases of the model, along with thermodynamic observables
and the space and imaginary time dependence of correlation functions.
The finite connectivity of the Bethe lattice induces a richer physical 
content with respect to its infinitely connected counterpart: 
a notion of distance between sites of the lattice is preserved, and the 
bosons are still weakly mobile in the Mott insulator phase.
The Bethe lattice construction can be viewed as an
approximation to the finite dimensional 
version of the model. We show indeed a quantitatively reasonable agreement
between our predictions and the results of Quantum Monte Carlo simulations in 
two and three dimensions.

\end{abstract}

\maketitle

\section{Introduction}

The bosonic version of the Hubbard model was introduced for the first time
in the seminal paper~\cite{FiFi}. It describes a system of bosons hopping
between neighboring sites of a lattice, and subjected to a local repulsive
interaction disfavouring the multiple occupancy of a site. The competition
between these two effects leads to a quantum phase transition~\cite{Sachdev} 
at zero temperature: when the hopping term dominates the groundstate is a
Bose-Einstein Condensate (BEC) delocalized over the lattice, also known as
a superfluid (SF) phase. On the contrary
for large local repulsion it becomes energetically favorable to form a
commensurate state where the average 
number of bosons per site is fixed to an integer
value. This incompressible phase is called a Mott insulator (MI).

The introduction of this model was motivated by experimental results on Helium
adsorbed on disordered substrates.
The recent progresses of the experimental techniques
for the manipulation of cold atoms, and in particular the possibility of
devising optical lattices loaded with cold gases, revivified the interest for
the Bose-Hubbard model. Following the proposal made in~\cite{bh_proposal_exp},
the experimental observation of this Mott transition was first achieved
in~\cite{bh_exp}. We refer the reader to~\cite{review_exp} for a review of such
experiments bridging a gap between atomic and condensed-matter physics.

From a more theoretical point of view the Bose-Hubbard model has been studied
with various techniques, notably mean-field 
approximations~\cite{FiFi,Gutzwiller},
perturbative expansions in the hopping 
strength~\cite{bh_perturb,bh_perturb2,freericks1},
Random Phase Approximations (RPA) \cite{Kampf93,Sheshadri93,vanOosten01,Sengupta05,Menotti08},
and Quantum Monte Carlo simulations~\cite{bh_num,bs_num,bh_qmc_2d,bh_qmc_3d,bh_qmc_number,hohenadler}. 
The mean-field approach of~\cite{FiFi} yields a qualitatively correct prediction
of the phase diagram of the model. However its description of the Mott
insulator is oversimplified, the hopping being completely neglected in this
phase at the mean-field level. As often the assumptions underlying the
mean-field approximation become valid when the connectivity of each site (the
number of its neighbors) is very large, either of the order of the system size
itself (in which case a rigorous treatment of the model is
possible~\cite{bh_rig_1,bh_rig_2}), or in the limit of large dimensionality.
The latter case was recently investigated in~\cite{BDMFT_1,BDMFT_2} where
a bosonic version of the Dynamic Mean-Field Theory~\cite{DMFT} was developed,
including first order corrections in the inverse of the spatial dimension.

In this paper we follow a related but slightly different road to improve on
the mean-field treatment of~\cite{FiFi}, by treating the Bose-Hubbard model
at the level of the Bethe approximation, i.e. by solving it on a Bethe lattice.
We show that, {\it in the thermodynamic limit},
the computation of all observables amounts to solve a single
self-consistent functional equation.
By writing this equation in the coherent state basis one recovers 
the B-DMFT in the large connectivity limit~\cite{BDMFT_1,BDMFT_2}.
Unfortunately, this equation cannot be analytically solved for finite
connectivity, except in some special limits.
Hence, we show how to rewrite it in a more convenient way using the
occupation number basis; this opens the way to its resolution through
an efficient numerical algorithm.

The Bethe approximation is the next step in a hierarchy
of approximations for lattice systems known as the Cluster Variation 
Method~\cite{CVM};  it is in fact exact when the model considered is
defined on a Bethe lattice, that is a graph which is locally a tree (no short 
loops can be closed on it). Bethe lattices appear in the DMFT 
derivations~\cite{DMFT,BDMFT_1,BDMFT_2} as an intermediate step, before taking
the limit of large connectivity. On the contrary in the present paper
the connectivity of the Bethe lattice will be kept finite. From the point 
of view of the universality classes of critical exponents Bethe lattice 
models fall in the mean-field category, yet their finite connectivity makes
them richer than the (fully-connected) mean-field version of~\cite{FiFi}.
In the latter any site is adjacent to all others, whereas on a Bethe lattice
one has a well defined notion of distance between two sites, though it does 
not correspond to the usual Euclidean distance. Moreover the MI phase is 
non-trivial for the Bethe lattice version of the model, at variance with the
mean-field picture. Though
we only consider here the ordered version of the model a motivation for
studying the Bethe lattice is the possibility that it exhibits a Bose Glass
phase~\cite{FiFi,inguscio1} in presence of disorder, which cannot be described at
the simplest mean-field level. This hope is backed up by the relevance of
the Bethe lattices for the localization problem~\cite{localization_Bethe}.

The methodology we use has its roots in the extensive research effort
which took place over the last decade in the community studying statistical 
mechanics of disordered systems. Under the name of cavity method a set of 
techniques has been developed to solve classical spin models on Bethe lattices
(in the sense of sparse random graphs), with applications to spin 
glasses~\cite{MePa_Bethe} and to random optimization problems arising from
theoretical computer science~\cite{ksat_science}. An extensive 
presentation of this field can be found in the recent book~\cite{cavity_book}.
The cavity method has been recently extended from classical to quantum spin 
models in presence of a transverse field~\cite{qcav_scardi,qcav_knysh,qcav_our}.
We widen here the range of applicability of the method by including lattice
boson models.

The rest of the paper is organized as follows.
In Sec.~\ref{sec_definition} we define precisely the model studied and the
physical observables of interest, and we recall the usual mean-field
treatment and its qualitative physical predictions.
Sec.~\ref{sec_overview} is an overview conveying the main ideas of the
cavity method, starting for pedagogical reasons from the ferromagnetic Ising 
model before explaining its extension to quantum problems, and in particular
its connection with the B-DMFT of~\cite{BDMFT_1,BDMFT_2}.
Sec.~\ref{sec_qcav_number} contains the core of our contribution; we first
present in Sec.~\ref{sec_qcav_number_eq} the main equations describing the
Bose-Hubbard model on the Bethe lattice, and the principles of their numerical 
resolution. In Sec.~\ref{sec_results} we present our predictions for several
physical observables and compare the phase diagram we obtain with the
Quantum Monte Carlo results in two~\cite{bh_qmc_2d} and three~\cite{bh_qmc_3d} 
dimensions, as well as with the B-DMFT prediction of~\cite{BDMFT_2} in three
dimensions. For the convenience of the reader not interested in technical
aspects we postpone the detailed derivation of the equations 
to Sec.~\ref{sec_derivation}.
Finally we present our conclusions and perspectives for future work in 
Sec.~\ref{sec_conclu}.

\section{Definitions and reminder on the traditional mean-field
approach}
\label{sec_definition}

\subsection{Definition of the model and of the observables}

We shall consider the Bose-Hubbard model for spinless bosons
on a graph of $N$ vertices 
(sites) defined by the following Hamiltonian,
\beq
H = - J \sum_\ij ( a^\dag_i a_j + a^\dag_j a_i ) 
+ \frac{U}{2} \sum_{i=1}^N a^\dag_i a^\dag_i a_i a_i - 
\mu \sum_{i=1}^N a^\dag_i a_i  = H_K + H_L \ ,
\label{eq_def_bh}
\eeq
where the first sum runs over a subset of the $N(N-1)/2$ possible edges (links)
between pairs of sites, $J$ is the hopping amplitude between
neighboring sites, $a_i$ (resp. $a^\dag_i$) is the boson annihilation
(resp. creation) operator on site $i$, $\mu$ is the chemical potential fixing
the density 
of particles, and $U$ controls the strength of the on-site interaction between
particles. It is convenient to separate the kinetic term $H_K$ (proportional
to $J$) in the Hamiltonian, from the local term $H_L$ including the Hubbard
interaction and the chemical potential term. Note that the kinetic energy defined in
such a way will be negative: indeed, the discrete lattice
version of the kinetic energy would be given by $H_K + J \sum_i c_i a^\dag_i a_i$ where
$c_i$ is the connectivity of site $i$.

Although the method we will discuss in this paper allows in principle to compute
very general observables (such as multi-point imaginary time correlations), 
in the following we will be mainly interested in standard observables such as
the mean density, kinetic and on-site energy, condensate fraction, and Green 
functions. For the sake of clarity we recall now their definitions.
The partition function at temperature $T$ is defined by
\beq
Z = \Tr \big[ \, e^{-\beta H} \big] \ ,
\eeq
where $\beta=1/T$ (we set $k_B = 1$), the corresponding free energy is
$F= -T \log Z $; the free-energy per site is $f=F/N$. 
The thermodynamic average of an operator $O$ is defined as
\beq
\la O \ra = \frac{1}{Z} \Tr \big[ \, O \, e^{-\b H} \big] \ .
\eeq
In particular the average number of particles on site $i$ reads 
$\la n_i \ra = \la a^\dag_i a_i \ra$ and
the density of particles is $\rho = N^{-1} \sum_i \la n_i \ra$;
the kinetic (resp. local) energy per site is $e_K = \la H_K \ra/N$ 
(resp. $e_L = \la H_L \ra/N$). To define the order parameter for BEC, one possibility
that is convenient for our purposes is
to introduce a small perturbation to the Hamiltonian in the form 
$H_\e = H+\e \sum_i (a_i +a^\dag_i)$
and define
\beq
\la a \ra = \lim_{\e \to 0} \la a \ra_\e \ ,
\eeq
where $\la \bullet \ra_\e$ denotes the thermodynamic average in presence of the perturbation.

The imaginary-time evolution of an operator $O$ is given by $O(\t) = e^{\t H}
O e^{-\t H}$;
we then define the two-times correlation functions as~\cite{Negele}
\beq\label{eq_green_ar}
\begin{split}
G^i_>(\t) & = \la a_i(\t) a_i^\dag(0) \ra = \frac{1}{Z} \Tr \big[ e^{-(\b-\t) H} a_i e^{-\t H} a^\dag_i \big] \ , \\
G^i_<(\t) & = \la a^\dag_i(0) a_i(\t) \ra = \frac{1}{Z} \Tr \big[ e^{-(\b+\t) H} a^\dag_i e^{\t H} a_i \big] \ .
\end{split}\eeq
The Green function is defined, for $-\beta/2 \leq \t \leq \beta/2$, by
\beq\label{eq_green_def}
G^i(\t) = \theta(\t) G^i_>(\t)+\theta(-\t) G^i_<(\t) = \la T a_i(\t) a^\dag_i(0) \ra \ ,
\eeq
where $T$ is the usual time-ordering operation, which should not be confused 
with the temperature. 
Note that the cyclic property of the
trace implies that $G^i_<(\t) = G^i_>(\t+\beta)$, hence
\beq
G^i(\t) = \theta(\t) G^i_>(\t)+\theta(-\t) G^i_>(\beta + \t) \ ,
\eeq
and the knowledge of $G^i_>(\t)$ for $0 \leq \t \leq \b$ is enough to determine
the Green function.

The cavity method allows also the computation of spatial correlation 
functions, we shall in particular determine the one-particle density matrix
$\rho_{ij}=\la a^\dag_i a_j \ra$.

\subsection{Review of mean-field results}
\label{sec_rev_MF}

The inexistence of an analytical solution of the model for finite 
dimensional lattices has triggered a large amount of research effort on
numerical simulations~\cite{bh_num,bs_num,bh_qmc_2d,bh_qmc_3d,bh_qmc_number}, 
pertubative expansions~\cite{bh_perturb,bh_perturb2}, 
or mean-field treatments~\cite{FiFi,Gutzwiller} of the problem.
Let us briefly recall the 
various points of view and methodology that the mean-field approach usually
encompasses. It can first be seen as an approximation to finite-dimensional
models. Maybe the most satisfactory way to perform this approximation is
by means of a variational method. One introduces a simpler trial (Gutzwiller)
Hamiltonian where the sites are decoupled,
\beq
H_0 = \sum_{i=1}^N \left[- \psi_i a^\dag_i - \psi_i^* a_i +  
\frac{U}{2} a^\dag_i a^\dag_i a_i a_i - \mu a^\dag_i a_i \right] \ ,
\label{eq_H0}
\eeq
the $\psi_i$ being here free parameters. Note that $H_0$ breaks the 
particle conservation symmetry. The free-energy at inverse
temperature $\beta$ of the original model, $F(\beta)$, can be upper-bounded as
$F \le F_0 + \langle H \rangle_0 - \langle H_0 \rangle_0$, where
$F_0(\beta)$ is the free-energy of the trial Hamiltonian and
$\langle \cdot \rangle_0$ denotes a thermal average with respect to the
trial Hamiltonian. The best variational description is thus obtained by 
minimizing the upper bound with respect to the parameters $\psi_i$. Assuming
that all sites of the graph have the same number $c$ of neighbors ($c=2d$ for 
a $d$-dimensional hypercubic lattice), the bound is minimized by taking the
same value $\psi$ on all sites, which can be chosen real without loss of
generality. The free-energy per site can then be upper-bounded as
$f(\beta) \le f_{\rm var}(\beta)$, with
\beq
f_{\rm var}(\beta) = \inf_{\psi} \left[
\frac{1}{c J} \psi^2 - \frac{1}{\beta} \ln \Tr \left[
e^{\beta \mu a^\dag a -\beta \frac{U}{2} a^\dag a^\dag a a + \beta \psi 
(a + a^\dag)}\right]
\right] \ ,
\label{eq_f_mf}
\eeq
where the trace is over the Hilbert space of a single site. The physical
properties of this variational free-energy are well-known~\cite{FiFi}: at
all temperatures and chemical potentials there is a transition encountered
upon increasing the hopping intensity $J$ from an ``insulating'' phase, 
characterized by $\langle a \rangle =0$, to a ``superfluid'' phase with 
$\langle a \rangle \neq 0$. At zero temperature this transition line draws a
series of lobes in the $(J/U,\mu/U)$ plane, inside each lobe the particle
density being constrained to a given integer. These Mott Insulator phases are
thus incompressible.
There exists alternative ways to obtain the estimation (\ref{eq_f_mf})
for the free-energy per site of the Bose-Hubbard Hamiltonian. The usual
mean-field approximation consists indeed in replacing
$a^\dag_i a_j$ with $ \langle a^\dag_i \rangle a_j  + 
a^\dag_i \langle a_j \rangle - \langle a^\dag_i \rangle \langle a_j \rangle$
in the hopping term of the original Hamiltonian, neglecting correlations 
between neighboring sites. This transforms the Hamiltonian into the
site-decoupled form (\ref{eq_H0}), with $\psi_i$ given by a sum of $\langle
a_j \rangle$ over the neighbors $j$ of $i$. These expectation values are then
computed with respect to the decoupled Hamiltonian, the self-consistency
equations leading finally to the same expression (\ref{eq_f_mf}) of the
free-energy per site as the variational approach; the latter has however the
advantage of being better controlled, in the sense that it provides a rigorous
bound on the true free-energy of the system. Finally, another reasoning
yielding this mean-field result consists in devising a model which has
exactly (\ref{eq_f_mf}) as its free-energy per site in the thermodynamic
limit, instead of taking it as an approximation for the finite-dimensional
case. As could be expected this model corresponds to the fully-connected
version of the Hamiltonian (\ref{eq_def_bh}), with the sum in the hopping term
running over all possible pairs of sites, with a coupling constant $J$
inversely proportional to the size $N$ of the system in order to have an
extensive thermodynamic limit. It has been shown rigorously
in~\cite{bh_rig_1,bh_rig_2} that in the thermodynamic limit the free-energy
of this fully-connected model converges indeed to (\ref{eq_f_mf}).

\begin{figure}
\begin{center}
\includegraphics[width=7cm]{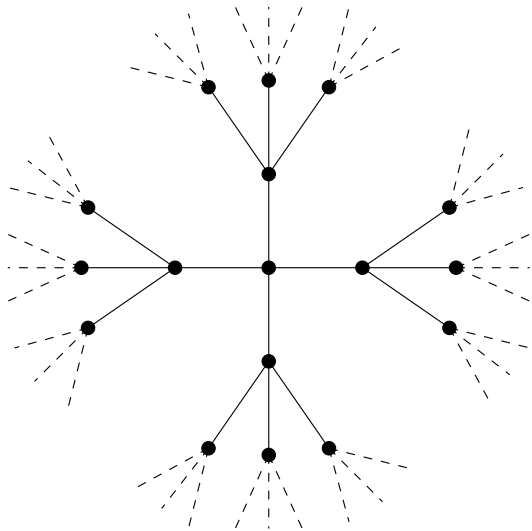}
\end{center}
\caption{A portion of a Bethe lattice of connectivity $c=4$, approximation
of a square lattice.}
\label{fig_Bethe}
\end{figure}

The above described mean-field treatment has limitations both of a
quantitative nature (the approximation cannot be expected to be very precise
for small dimensions) and of a qualitative one. In particular the MI
phase is rather trivial; as the infimum in (\ref{eq_f_mf}) is reached in
$\psi=0$, the hopping of particles is completely suppressed in this phase. 
This drawback is particularly severe in the case of the disordered
Bose-Hubbard model, for which it forbids the description of the Bose Glass
phase predicted in~\cite{FiFi}. In order to cure this defect of the mean-field
treatment one has to account in some way for the correlations between
neighboring sites. Thinking in terms of classical spin models, the mean-field
approximation is the lowest level of a hierarchy of descriptions (known as
Cluster Variation Method~\cite{CVM}, or Kikuchi 
approximations~\cite{Kikuchi}) which treats exactly
larger and larger regions of the interaction graph. In this work we shall
deal with the Bose-Hubbard model at the next level of the hierarchy, 
known as the Bethe approximation, 
in which correlations between nearest neighbors are
explicitly taken into account. In the same way as
the simplest mean-field approximation was
exact for the fully-connected model, the Bethe approximation is exact for a
family of graphs, known as Bethe lattices. In these graphs each site interacts
with a finite number $c$ of neighbors, say $2d$ in order to mimic a
$d$-dimensional hypercubic lattice, but the short loops present in
the latter are discarded: Bethe lattices have a local tree
structure, see Fig.~\ref{fig_Bethe} for a picture of the local appearance
of a Bethe lattice of connectivity $c=4$. 
For at least two reasons it is however better not to picture a
Bethe lattice as a finite regular tree (usually called Cayley tree in this
context). Indeed a regular tree is strongly inhomogenous, a finite fraction of
its ``volume'' being very close to its ``surface'', and only the center of the
Cayley tree has the bulk properties one is interested in. Moreover in the case
of frustrated models the boundary conditions imposed on the leaves of the
Cayley tree have to be chosen with particular care. For these two reasons it
has become customary in the community of statistical mechanics of disordered
systems, following~\cite{MePa_Bethe}, to define a Bethe lattice as a random
$c$-regular graph~\cite{rgraphs}, that is a graph chosen uniformly at random
from the set of the graphs on $N$ vertices where all sites have precisely
$c$ neighbors. These Bethe lattices are locally tree-like, their loops have
typically a length diverging with the size $N$ of the system, yet they do not 
have any boundary, all sites playing the same role (in the same way as
periodic boundary conditions impose translation invariance on a finite cubic 
lattice). The absence of an underlying
finite-dimensional structure gives them a mean-field character, but their
finite connectivity leads to a richer content than the fully-connected
models. The free-energy of Bethe lattice models is self-averaging with 
respect to their random character in the thermodynamic limit. In other words
for large enough $N$ a single sample is a good representative of the ensemble
average.
The so-called cavity method has been developed to treat classical spin
models defined on such random graphs, and found important applications for
optimization problems of theoretical computer science~\cite{ksat_science}.

\section{Overview of the quantum cavity method}
\label{sec_overview}

\subsection{The cavity method for ferromagnetic Ising models}
\label{sec_Ising}

For the sake of pedagogy we shall begin our presentation by the cavity method
treatment of the ferromagnetic Ising model on the Bethe lattice. We consider 
the following energy function,
\beq
H(\s_1,\dots,\s_N) = - J \sum_\ij \s_i \s_j - h \sum_i \s_i \ ,
\eeq
where the classical degrees of freedom $\s_i$ can take the values $\pm 1$,
the first term describing ferromagnetic ($J>0$) interactions between neighbors
on a graph of $N$ vertices, the second one corresponding to an uniform
magnetic field. We shall repeatedly use in the following the notation $\di$
for the set of vertices adjacent to a given vertex $i$, i.e. for the sites
which interact with $i$, and $\dimj$ for those vertices around $i$ distinct
from $j$. The Gibbs-Boltzmann probability measure is
\beq
\eta(\s_1,\dots,\s_N) = \frac{1}{Z} e^{-\beta H(\s_1,\dots,\s_N)} \ ,
\label{eq_law_Ising}
\eeq
with the partition function $Z$ ensuring its normalization. The goal is to
compute the free-energy per site $f=-(\ln Z)/(N\beta)$ and the local
magnetizations $\langle \s_i \rangle$, the brackets denoting an average with
respect to the law (\ref{eq_law_Ising}).

\begin{figure}
\begin{center}
\includegraphics{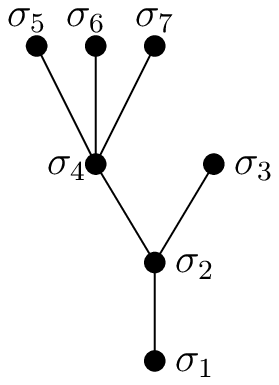}
\end{center}
\caption{Example of an Ising tree model on 7 vertices. The definition of
$Z_{2\to 1}$ and its recursive computation reads here:
$Z_{2\to 1}(\s_2)=\underset{\s_3,\dots,\s_7}{\sum} 
e^{\beta h (\s_2 + \s_3 + \dots + \s_7) + \beta J (\s_2 (\s_3 + \s_4) + \s_4
  (\s_5 + \s_6+\s_7)) }= e^{\beta h \s_2} \underset{\s_3,\s_4}{\sum} Z_{3 \to
  2}(\s_3) Z_{4 \to 2}(\s_4) e^{\beta J \s_2 (\s_3 +\s_4)} $. 
}
\label{fig_Ising_example}
\end{figure}

Let us first consider the case where the interaction graph is a finite
tree. In this case the computation can be organized in a very simple way,
taking benefit of the natural recursive structure of a tree. We define the
quantity $Z_{i \to j}(\s_i)$, for two adjacent sites $i$ and $j$, as the 
partial partition function for the subtree rooted at $i$, excluding the branch
directed towards $j$, with a fixed value of the spin variable on the site
$i$. We also introduce $Z_i(\s_i)$, the partition function of the whole tree
with a fixed value of $\s_i$. These quantities can be computed according to
the following recursion rules, see Fig.~\ref{fig_Ising_example} for an
example, 
\beq
Z_{i \to j}(\s_i) = e^{\beta h \s_i} 
\prod_{k \in \dimj} \left(\sum_{\s_k} Z_{k \to i}(\s_k) e^{\beta J \s_i \s_k} 
\right) \ , \qquad
Z_i(\s_i) = e^{\beta h \s_i} \prod_{j \in \di} \left( 
\sum_{\s_j} Z_{j \to i}(\s_j) e^{\beta J \s_i \s_j} \right) \ .
\eeq
It will be useful for the following discussion to rewrite these equations in
terms of normalized quantities which can be interpreted as probability laws
for the random variable $\s_i$, namely 
$\eta_{i \to j}(\s_i)=Z_{i \to j}(\s_i)/\sum_{\s'}Z_{i \to j}(\s')$ and
$\eta_i(\s_i)=Z_i(\s_i)/\sum_{\s'}Z_i(\s')$. The recursion equations read in
these notations
\beq
\eta_{i \to j}(\s_i) = \frac{1}{z_{i \to j}} e^{\beta h \s_i} 
\prod_{k \in \dimj} \left(\sum_{\s_k} \eta_{k \to i}(\s_k) 
e^{\beta J \s_i \s_k} \right) \ , \qquad
\eta_i(\s_i) = \frac{1}{z_i} e^{\beta h \s_i} \prod_{j \in \di} \left( 
\sum_{\s_j} \eta_{j \to i}(\s_j) e^{\beta J \s_i \s_j} \right) \ ,
\label{eq_msg_Ising}
\eeq
where $z_{i \to j}$ and $z_i$ are normalization constants.
On a given tree the recursion equations on the $\eta_{i \to j}$ for all
directed edges of the graph have a single solution, easily found by
propagating the recursion from the leaves of the graph. Moreover the quantity
$\eta_i(\s_i)$ is exactly the marginal probability law of the Gibbs-Boltzmann
distribution (\ref{eq_law_Ising}), hence the local magnetizations can be
computed as $\langle \s_i \rangle = \sum_{\s} \eta_i(\s) \s$.

\begin{figure}
\begin{center}
\includegraphics{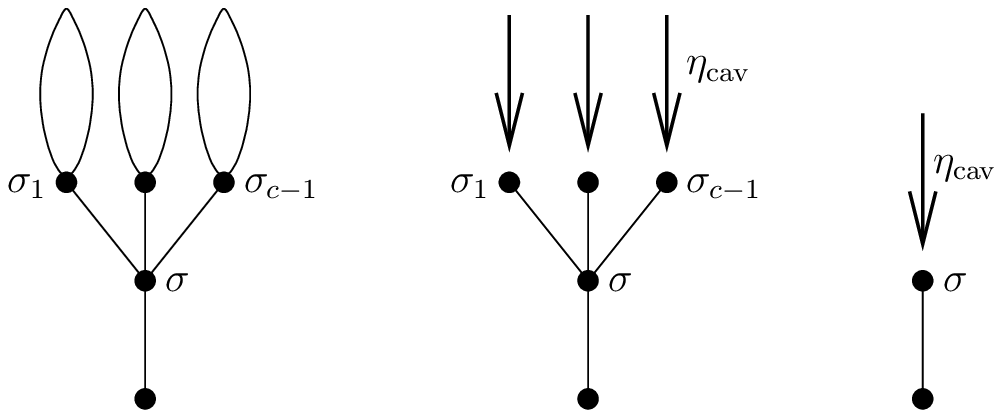}
\end{center}
\caption{Pictorial representation of Eq.~(\ref{eq_fpoint_Ising}). The bubbles
on the first panel represent subtrees of the graph; their effect on the spins
$\s_1,\dots\s_{c-1}$ is summarized by $\eta_{\rm cav}$, represented as a bold
arrow in the second panel. Tracing over these $c-1$ spins leads to the third
panel.}
\label{fig_Ising_iter}
\end{figure}

\begin{figure}
\begin{center}
\includegraphics{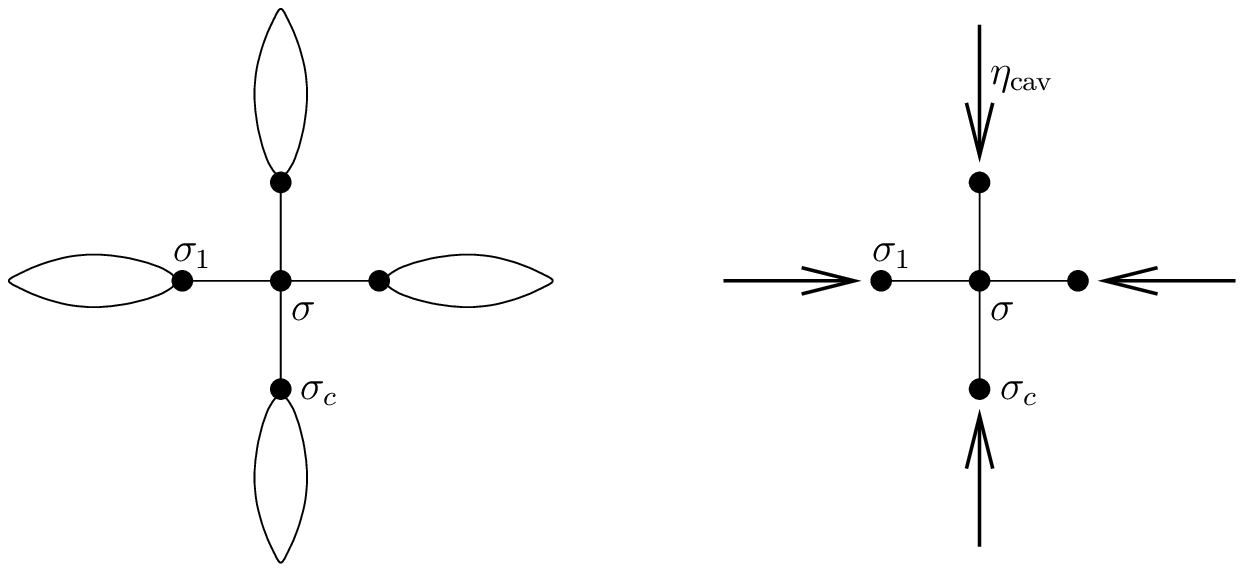}
\end{center}
\caption{Illustration of Eq.~(\ref{eq_Ising_final}); the local magnetization
  of one site is computed by taking into account all the $c$ neighbors.} 
\label{fig_Ising_final}
\end{figure}

The reasoning above was made under the assumption that the interaction graph
was a tree. One can however look for a solution of the recursion equations
(\ref{eq_msg_Ising}) on any graph, even in the presence of loops. This
approach is known as Belief Propagation in inference problems~\cite{fgraphs}, 
and corresponds
to the Bethe approximation of statistical mechanics~\cite{Yedidia}. The
cavity method is a set of prescriptions to use these local recursion equations
to predict the behavior of models defined on sparse random graphs, that are
locally tree-like (the typical length of the loops diverging in the
thermodynamic limit). In its simplest version (known as the replica symmetric
one) one makes the assumption of the existence of a single pure state 
which implies that the effect of the loops
does not spoil the local recursions (\ref{eq_msg_Ising}). Their presence
simply provides self-consistent boundary conditions and avoids the ill-defined
behaviour due to the dominant surface effects of trees.
In the case of unfrustrated ferromagnetic models it has
been shown rigorously~\cite{ferro_rigorous} that this assumption is correct,
the predictions of the cavity method being exact in the thermodynamic limit,
both for the local magnetizations and for the free-energy per site. Let us
be more explicit for the case of the Bethe lattice, where all vertices have
the same connectivity $c$. The probability laws $\eta_{i \to j}$ are then
the same on all edges; denoting $\eta_{\rm cav}$ their common value, one
finds the self-consistent equation
\beq
\eta_{\rm cav}(\s) = \frac{1}{z_{\rm cav}}e^{\beta h \s} 
\sum_{\s_1,\dots,\s_{c-1}} 
\eta_{\rm cav}(\s_1)\dots \eta_{\rm cav}(\s_{c-1}) 
e^{\beta J \s(\s_1 + \dots + \s_{c-1} )} \ ,
\label{eq_fpoint_Ising}
\eeq
with $z_{\rm cav}$ a normalization constant. A pictorial representation of
this equation can be found in Fig.~\ref{fig_Ising_iter}. The local
magnetization is then computed as
\beq
\langle \s \rangle = \sum_\s \eta(\s) \s \ , \qquad
\eta(\s) = \frac{1}{z} e^{\beta h \s} 
\sum_{\s_1,\dots,\s_c} \eta_{\rm cav}(\s_1)\dots \eta_{\rm cav}(\s_c) 
e^{\beta J \s(\s_1 + \dots + \s_c)} \ ,
\label{eq_Ising_final}
\eeq
including the $c$ neighbors of a central site as represented in
Fig.~\ref{fig_Ising_final}. The term cavity comes from the fact that
in Eq.~(\ref{eq_fpoint_Ising}) one site has been removed from the neighborhood
of the considered vertex. As the Ising variable $\s$ can only take two values,
each of the probability distributions $\eta_{\rm cav}$ and $\eta$ can be
parametrized by a single real number,
a cavity (resp. effective) magnetic field,
\beq
\eta_{\rm cav}(\s) = 
\frac{e^{\beta h_{\rm cav} \s}}{2 \cosh(\beta h_{\rm cav})} \ , \qquad
\eta(\s)=\frac{e^{\beta h_{\rm eff} \s}}{2 \cosh(\beta h_{\rm eff})} \ ,
\eeq
solutions of
\beq
h_{\rm cav} = h + \frac{c-1}{\beta} \text{arctanh}[\tanh(\beta J) \tanh(\beta
h_{\rm cav} ) ] \ , \qquad
h_{\rm eff} = h + \frac{c}{\beta} \text{arctanh}[\tanh(\beta J) \tanh(\beta
h_{\rm cav} ) ] \ .
\eeq
thus making the resolution of 
(\ref{eq_fpoint_Ising}),(\ref{eq_Ising_final}) extremely simple.
In particular at zero external field $h=0$, 
one finds that a phase transition occurs at $\beta = \beta_c$,
separating a high temperature paramagnetic phase ($h_{\rm cav}=h_{\rm eff}=0$)
from a low temperature ferromagnetic phase ($h_{\rm cav}, h_{\rm eff} \neq 0$).
The critical temperature is easily obtained linearizing the equation for $h_{\rm cav}$
around $h_{\rm cav}=0$ and is the solution of $(c-1) \tanh(\b_c J) = 1$.

It is worth noting that one can compute explicitly the spin-spin correlation
function, which is given in the paramagnetic phase 
by $C_{ij} = \la \s_i \s_j \ra = [ \tanh(\b J) ]^{d(i,j)}$, where $d(i,j)$ is
the distance between sites $i$ and $j$, defined as the length of the shortest path
connecting sites $i$ and $j$. The associated correlation
length is $\xi = -\log[\tanh(\b J)]$ which 
is finite at the transition point $\b = \b_c$.
Nevertheless, 
the associated magnetic susceptibility $\chi = N^{-1}\sum_i \dd \la \s_i \ra / \dd h$ 
diverges: one can show using the fluctuation-dissipation theorem
that $\chi = \b N^{-1} \sum_{ij} C_{ij} =\b \sum_{d=0}^\io \NN_d C_d$, where $\NN_d$ is the
number of points at distance $d$ from a given reference point and scales as $(c-1)^d$ for
large $d$. Therefore, if $C_d$
decays slower than $(c-1)^{-d}$, the corresponding susceptibility is divergent; this is
indeed consistent with the equation for $\b_c$ obtained above.
Hence phase transitions on Bethe lattices are always associated to diverging
susceptibilities and finite correlations lengths (see~\cite{Farhi} for a
discussion of this fact in the context of quantum spin models). 
For finite dimensional
lattices, $\NN_d$ grows as a power of $d$, and one needs a diverging correlation
length to obtain a diverging susceptibility.

\subsection{Suzuki-Trotter formalism and quantum cavity method}
\label{sec_coherent}

We have just seen how the cavity method deals with classical spin models on
locally tree-like graphs. The extension of the method to quantum models is
based on the Suzuki-Trotter formula. Generically speaking the resolution of a
quantum model amounts to compute its partition function 
$Z=\Tr \left[e^{-\beta H}\right]$, 
where the Hamiltonian $H$ is now an operator, which can usually be split as 
$H=H_1 + H_2$ with $H_{1,2}$ two operators which do not commute. For instance
in the case of quantum spin 1/2 models one can have an interaction term $H_1$
involving only the $z$ component of the spins, 
while $H_2$ is a transverse field acting in a perpendicular direction. 
In the case of the Bose-Hubbard model this
decomposition splits the Hamiltonian between the hopping term $H_K$
and the on-site potential energy $H_L$. 
The non-commutativity of $H_{1,2}$ can be cured with the
application of the Suzuki-Trotter formula~\cite{suzuki},
\beq
Z = \lim_{\Ns \to \infty} \Tr \left[\left(e^{-\frac{\beta}{\Ns}H_1} 
e^{-\frac{\beta}{\Ns}H_2} \right)^{\Ns} \right] \ .
\label{eq_Suzuki}
\eeq
The computation then proceeds with the insertion of $\Ns$ representations of 
the identity between the $\Ns$ elements of this product. It is convenient to
express the identity operator in a basis where one of the two operators is
easily evaluated. In the case of quantum spin 1/2 models one can for instance
use the eigenvectors of the Pauli matrix in the $z$ direction; 
the quantum partition function is then
turned into the partition function of a classical Ising spin model, where each
quantum spin is replaced by a set of $\Ns$ classical spins, indexed by their
position on the original graph and a supplementary ``discrete imaginary time''
coordinate (which becomes a continuous parameter in the $\Ns\to\infty$
limit). The important point is that the spatial structure of the graph of
interactions is preserved in this construction, at the price of the
introduction of imaginary time-dependent classical degrees of freedom.
In particular, if the interactions of the quantum model are defined on a
tree-like graph, the cavity method still applies to this extended classical 
model. This line of thought was first followed for quantum spin models 
in~\cite{qcav_scardi} (see also~\cite{qcav_knysh}) for a finite number $\Ns$
of Suzuki-Trotter slices, the continuous imaginary time limit was then taken 
in~\cite{qcav_our}. In this paper we shall extend this method to deal with
lattice bosons models.

In this case the decomposition of the identity operator in the Suzuki-Trotter
can be expressed using either coherent states or occupation numbers.
The latter has the advantage of being discrete,
and we shall use it in the rest of the paper. For the sake of clarity and in
order to make contact with the recently proposed B-DMFT~\cite{BDMFT_1,BDMFT_2}
we discuss first the application of the cavity method within the coherent states 
representation in the rest of this subsection. Inserting such a decomposition
of the identity for each of the $N$ sites at each of the $\Ns$ Suzuki-Trotter
slices leads, in the continuous time limit $\Ns \to \infty$, to the coherent
state path integral expression of the partition function of the Bose-Hubbard
model~\cite{Negele}: 
\bea
Z &=& \int \prod_{i=1}^N D\ba_i  D\bac_i \, e^{-S} \ , \\
S &=& \int_0^\beta \dd \t \, \left[
\sum_{i=1}^N \left( \ac_i(\t) (\partial_\t - \mu ) a_i(\t) + \frac{U}{2}
  (\ac_i(\t) a_i(\t))^2 \right)
-J \sum_\ij \left( \ac_i(\t) a_j(\t) + \ac_j(\t) a_i(\t) \right) \right] \ .
\eea
Here and in the following we use bold symbols to denote imaginary-time
dependent quantities. $a_i(\t)$ and $\ac_i(\t)$ are two (formally conjugate)
complex numbers indexing the coherent state at site $i$ and imaginary time
$\t$, with $D\ba_i  D\bac_i$ the path-integral measure of this site. Following
the same steps as in the study of the Ising model, the cavity method for a
Bethe lattice of connectivity $c$ leads to the following self-consistency
equation: 
\beq
\eta_{\rm cav}(\ba,\bac) = \frac{1}{z_{\rm cav}} w(\ba,\bac)
\int \prod_{i=1}^{c-1} D\ba_i D\bac_i \, \eta_{\rm cav}(\ba_i,\bac_i) \
\exp \left\{ J \int_0^\b \dd \t \left[ 
\ac(\t) \sum_{i=1}^{c-1} a_i(\t) + a(\t) \sum_{i=1}^{c-1} \ac_i(\t) \right]
\right\} \ ,
\label{eq_fpoint_coherent}
\eeq
with the on-site weight of the path $(\ba,\bac)$ given by
\beq
w(\ba,\bac) = \exp \left[-\int_0^\b \dd \t \left( 
\ac(\t) (\partial_\t - \mu) a(\t) + 
\frac{U}{2} (\ac(\t) a(\t))^2 \right)\right] \ .
\eeq
This self-consistent equation on $\eta_{\rm cav}$ is formally similar to the
corresponding Eq.~(\ref{eq_fpoint_Ising}) of the Ising model. It is however
much more complicated: the Ising degree of freedom $\s$ could only take two
distinct values, whereas $(\ba,\bac)$ belongs to a space of functions of the
imaginary time $\t$. 
Therefore $\h_{\rm cav}(\ba,\bac)$ is a functional measure whose representation 
is much more difficult; a complete parametrization requires the knowledge of all $n,m$-points
correlations $\la a(t_1) \cdots a(t_n) a^\dag(s_1) \cdots a^\dag(s_m) \ra$.
On the other hand, this is one of the most interesting features of the quantum cavity method:
on-site quantum fluctuations are fully kept into account, without any truncation of higher order
correlations.

Unfortunately, an exact solution of (\ref{eq_fpoint_coherent}) can be
easily obtained only in the case of free bosons ($U=0$). $\eta_{\rm cav}$ 
acquires in this case a Gaussian form, with averages and two-point functions 
which can be
computed exactly and reproduce the results obtained by direct
diagonalization of the adjacency matrix of the Bethe
lattice~\cite{spectrum_Bethe}. 

\subsection{Large connectivity limit and the connection with B-DMFT}
\label{sec_BDMFT}

In the interacting case ($U>0$) a solution of 
(\ref{eq_fpoint_coherent}) can be looked for in the limit of large
connectivity, and this is precisely the road followed by the B-DMFT
studies~\cite{BDMFT_1,BDMFT_2}. For completeness we shall explain in this 
subsection how to recover the B-DMFT formalism from 
Eq.~(\ref{eq_fpoint_coherent}), before turning in the next section to
the occupation number basis which will allow to solve the model
for any connectivity.

To state the large $c$ expansion let us
rewrite (\ref{eq_fpoint_coherent}) in a more convenient way, using
the lighter Nambu notation
$\Psi^\dag(\t) = (\ac(\t),a(\t))$, and consequently $\bPsi^\dag = (\bac,\ba)$.
Then we can rewrite (\ref{eq_fpoint_coherent}) as
\beq\label{eq_fpoint_coherent_simple}
\eta_{\rm cav}(\bPsi) = \frac{1}{z_{\rm cav}} \, w(\bPsi) \, 
e^{(c-1) \G(J\bPsi)} \ , 
\qquad \G(\bPhi) = \log \left\{\int D\bPsi \, \eta_{\rm cav}(\bPsi)
\exp \left[\int_0^\beta \dd \t \ \Psi^\dag(\t) \Phi(\t) \right]\right\} \ .
\eeq
$\G(\bPhi)$ is the generating functional of the connected correlation 
functions of $a$ and $\ac$.
It can be expanded as
\beq\label{cumu_exp}
\G(\bPhi) = \int_0^\beta \dd \t \, \la \Psi^\dag \ra \Phi(\t) + 
 \int_0^\beta \dd \t \dd \t'\, 
\Phi^\dag(\t) \wh G_c(\t-\t') \Phi(\t') + O(\Phi^3) \ ,
\eeq
where the averages $\la \cdot \ra$ are with respect to $\eta_{\rm cav}$, we
have used the cyclic invariance in imaginary time, and 
$\wh G_c(\t-\t') = \la \Psi(\t) \Psi^\dag(\t') \ra - \la \Psi \ra \la
\Psi^\dag \ra$ is the connected part of the two point correlator of 
$\bPsi$. 

In the large connectivity limit, the superfluid-insulator transition happens
for a critical value of $J = \JJ/c$, with a finite $\JJ$. This can be argued by
looking at Eq.~(\ref{eq_f_mf}), where it is clear that the dependence on
hopping and 
connectivity is only through $\JJ = J c$, which is the real control parameter.
For large $c$ and $J = \JJ/c$, the cumulant expansion of 
$(c-1) \G(\JJ \bPsi/c)$ thus becomes a systematic expansion in powers
of $1/c$. By keeping only a finite number of terms in the cumulant
expansion, we obtain an expression of $\eta_{\rm cav}(\bPsi)$ that 
{\it is not} Gaussian (because of the $U$ term in $w(\bPsi)$); still, we can
obtain closed equations for the cumulants by computing them self-consistently 
as averages over the non-Gaussian $\eta_{\rm cav}$.

The leading order in $c$ gives, assuming without loss of generality 
that the average value of the order parameter is real,
$\la \Psi^\dag \ra = (\psi,\psi)$,
\beq\label{eq_coherent_leading}
\eta_{\rm cav}(\bPsi) = \frac{1}{z_{\rm cav}} \, w(\bPsi) \, 
e^{\int_0^\beta \dd \t \, \JJ \la \Psi^\dag \ra \Psi(\t)} \hskip10pt
\Rightarrow \hskip10pt
\psi =  
\frac{1}{z_{\rm cav}} \int D\ba D\bac \, a(0) \,
e^{-\int_0^\b \dd \t \left( 
\ac(\t) (\partial_\t - \mu) a(\t) + \frac{U}{2} (\ac(\t) a(\t))^2 
-\JJ \psi ( a(\t) + \ac(\t) )  \right)}  \ .
\eeq
This last equation can be rewritten in the operator representation, which
gives back the equation for $\psi$ corresponding to the minimization
of the variational free-energy (\ref{eq_f_mf}), up to
a multiplicative constant in the definition of $\psi$.
Note that a generalization of the discussion above and
of Eq.~(\ref{eq_coherent_leading}) to the disordered case leads to the
stochastic mean field theory devised in \cite{SMFT}.

We will now show that the next-to-leading order in the cumulant expansion of
(\ref{cumu_exp}) gives the B-DMFT
equations recently derived in~\cite{BDMFT_1,BDMFT_2}. Note that the
truncation at this two-point level was also used in the
context of spin models in~\cite{qcav_scardi}.
Plugging the expansion (\ref{cumu_exp}) in (\ref{eq_fpoint_coherent_simple}), 
we obtain to order $1/c$:
\beq\begin{split}
&\eta_{\rm cav}(\bPsi) = \frac{1}{z_{\rm cav}} \, \exp \big[ - S_{\rm loc} ] 
\ , \\
&S_{\rm loc} = 
 \int_0^\beta \dd \t \dd \t'\,
\Psi^\dag(\t) \wh \GG^{-1}(\t-\t') \Psi(\t')
+ \int_0^\beta \dd \t \, \left[
\frac{U}8 (\Psi^\dag(\t) \Psi(\t) )^2 -
\JJ \frac{c-1}c \la \Psi^\dag \ra \Psi(\t) 
\right] \ , \\
&\wh \GG^{-1}(\t-\t') = \frac12 \left( \begin{matrix} \partial_\t  -\mu & 0 \\
0 & -\partial_\t - \mu \end{matrix} \right) \d(\t-\t') - 
\frac{\JJ^2}c \wh G_c(\t-\t') \ .
\end{split}\eeq
Then, $\la \Psi^\dag \ra$ and 
$\wh G_c(\t-\t')=\la \Psi(\t) \Psi^\dag(\t') \ra - \la \Psi \ra \la \Psi^\dag
\ra$  have to be computed self-consistently as averages with the local action
$S_{\rm loc}$. This set of equations correspond exactly to the B-DMFT
of~\cite{BDMFT_1,BDMFT_2} for the special case of a Bethe lattice.

Away from these two limits ($U=0$ and $c \to \infty$) it seems difficult to
obtain a solution of the cavity equation (\ref{eq_fpoint_coherent}) as written
in the coherent state basis. As a consequence we shall turn in the following to
the representation number basis to apply the Suzuki-Trotter formula and thus
obtain a more tractable equation for all values of $U$ and $c$.

\section{The quantum cavity method in the occupation number basis}
\label{sec_qcav_number}
\subsection{The equations and the procedure for their numerical resolution}
\label{sec_qcav_number_eq}

The insertion of a decomposition of the identity expressed in the occupation number
basis in the Suzuki-Trotter formula (\ref{eq_Suzuki}) leads to an
expression of the partition function of the Bose-Hubbard model as a sum over
occupation number trajectories in imaginary time, $\{n_i(\t)\}$. These
trajectories are defined on an imaginary time interval of length $\beta$, with
the periodicity condition $n_i(0)=n_i(\beta)$. The weight 
(action) of these trajectories has two origins: the local part of the
Hamiltonian (\ref{eq_def_bh}) yields a contribution of the form 
$\exp[-\int \dd \t \, V(n_i(\t))]$ 
for each of the sites, where $V(n) = U n (n-1)/2 - \m n$
is the local energy term in the Hamiltonian.
In addition the hopping 
term of the Hamiltonian imposes constraints between the occupation number
trajectories: each time $n_i(\t)$ is raised (resp. decreased) by 1, the
occupation number $n_j(\t)$ of one of the neighbors $j \in \di$ must decrease
(resp. increase) of 1, meaning one particle has jumped from $j$ to $i$
(resp. from $i$ to $j$). Moreover each hopping event multiplies the weight of
the trajectory $\{n_i(\t)\}$ by $J$ and by a coefficient depending on the
instantaneous occupation numbers of the sites involved in the hopping. An
explicit derivation of this representation shall be given in
Sec.~\ref{sec_derivation}. Here we wish to present the results of the
computation in a lighter way for the ease of the reader not interested in the
technical details (we follow the methodology developed for quantum spin models
in~\cite{qcav_our}).

\begin{figure}
\includegraphics[width=8cm]{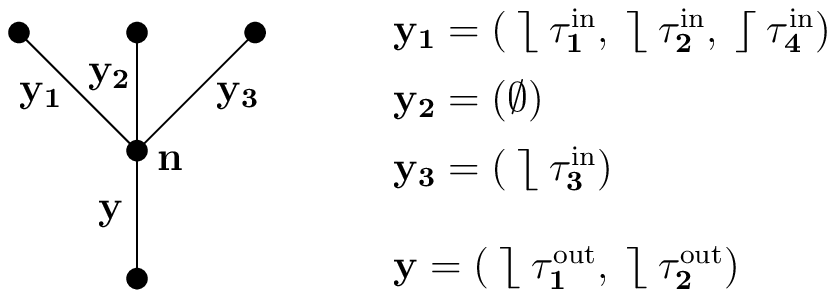}
\includegraphics[width=8cm]{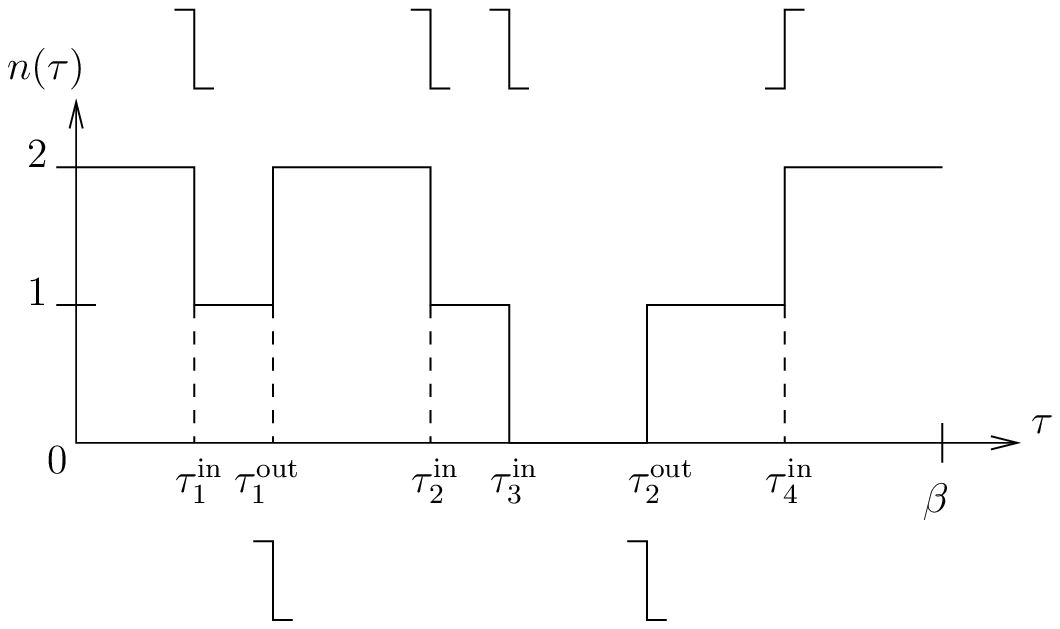}
\caption{
Example of the sampling process from the law
  $P(\by|\by_1,\dots,\by_{c-1})$ defined in Eq.~(\ref{eq_PZ}) for $c=4$. 
The $c-1$ incoming hopping trajectories $(\by_1,\by_2,\by_3)$ impose jumps in 
$\t_1^{\rm in}<\t_2^{\rm in}<\t_3^{\rm in}<\t_4^{\rm in}$, a particle arriving
  on the central site at the first three imaginary times, leaving at the
  fourth one, corresponding to the symbols in the upper part of the rightmost figure. A
  trajectory $n(\t)$ respecting these constraints is shown; it contains two
  other jumps at $\t_1^{\rm out} < \t_2^{\rm out}$, associated to hopping
  events in $\by$. The symbols on the bottom are reversed: particles leaving
  the central site will arrive on the next level of the recursive equation.} 
\label{fig_traj_iter}
\end{figure}

The recursion equations of the Bose-Hubbard model defined on a tree (or on a
locally tree-like graph with the assumptions of the cavity method) can be
expressed in terms of the probability distribution $\eta_{\rm cav}(\by)$ of the
hopping trajectories $\by$ defined on the edges of the graph. 
This quantity $\by$ encodes the imaginary times at which a particle has crossed
the edge, and the directions of the jumps; 
examples are given in Fig.~\ref{fig_traj_iter}. 
In the case of a regular Bethe
lattice of connectivity $c$ one obtains the following self-consistent equation 
on $\eta_{\rm cav}(\by)$,
\beq
\eta_{\rm cav}(\by) = \frac{1}{z_{\rm cav}} w_{\rm link}(\by) 
\sum_{\by_1,\dots,\by_{c-1}} \eta_{\rm cav}(\by_1) \dots 
\eta_{\rm cav}(\by_{c-1}) \,  w_{\rm iter}(\by,\by_1,\dots,\by_{c-1}) \ .
\label{eq_fpoint_number}
\eeq
For simplicity
we denote here with a sum symbol what is actually an integral over continuous
degrees of freedom $\by$.
The explicit expressions for the weights $w$ shall be given in
Sec.~\ref{sec_derivation}, see Eqs.~(\ref{eq_wlink}), (\ref{eq_witer}). Let us emphasize the formal similarity with 
the Ising equivalent Eq.~(\ref{eq_fpoint_Ising}), and of course the greater
complexity of the basic degree of freedom in the Bose-Hubbard case, the
hopping trajectory $\by$ assuming values in a much larger space than the Ising
variable $\s \in \{+1,-1\}$. We already mentioned this problem while
discussing the related equation (\ref{eq_fpoint_coherent}) in the coherent
state basis. Compared to this latter case we are however facing now an easier
problem: the number of hopping events on a given edge is a (random) number
which remains finite as long as the temperature is positive (it is actually  
of order $\beta J$). A single hopping trajectory $\by$
can thus be encoded as an integer $p$, i.e. the number of particle jumps
occuring on this edge during the imaginary time interval $[0,\beta]$, 
$p$ reals precising the imaginary times where these
jumps occur, and $p$ binary variables giving the direction of the jumps, 
see Fig.~\ref{fig_traj_iter}. In
contrast the coherent state trajectories were those of two continuous
functions with a priori no compact representation. This remark opens the way
to an efficient numerical method for the resolution of
(\ref{eq_fpoint_number})~\footnote{the code of our implementation is available
upon request.}.
We can indeed follow the population dynamics 
strategy~\cite{localization_Bethe,MePa_Bethe}. 
The idea of this method is to represent numerically
the probability distribution $\eta_{\rm cav}(\by)$ as a (weighted) sample of
a large number $\Nt$ of hopping trajectories, namely
\beq
\eta_{\rm cav}(\by) = \sum_{i=1}^\Nt g_i \, \delta(\by - \by_i ) \ ,
\label{eq_rep_popu}
\eeq
where the $\Nt$ weights of the trajectories are normalized according to
\beq 
\sum_{i=1}^\Nt g_i = 1 \ .
\label{eq_norm}
\eeq
Sampling an element $\by$ from the probability distribution $\eta_{\rm cav}$ 
corresponds in this representation to extract an integer $i\in[1,\Nt]$ with
probability $g_i$, and setting $\by= \by_i$. This representation of 
$\eta_{\rm cav}$ is an approximation, which yields better and better 
numerical accuracy when $\Nt$ grows. The determination of a sample of weights
$g_i$ and trajectories $\by_i$ which turns the representation 
(\ref{eq_rep_popu}) into a good approximation of the solution of 
(\ref{eq_fpoint_number}) can be performed iteratively. To explain this point
let us first rewrite the self-consistent equation (\ref{eq_fpoint_number}) as
\beq
\eta_{\rm cav}(\by) = 
\sum_{\by_1,\dots,\by_{c-1}} \eta_{\rm cav}(\by_1) \dots 
\eta_{\rm cav}(\by_{c-1}) P(\by|\by_1,\dots,\by_{c-1} )
\frac{\Z(\by_1,\dots,\by_{c-1})}{z_{\rm cav}} \ ,
\label{eq_fpoint_number2}
\eeq
where we have defined
\beq
P(\by|\by_1,\dots,\by_{c-1} ) = 
\frac{w_{\rm link}(\by) w_{\rm iter}(\by,\by_1,\dots,\by_{c-1})}
{\Z(\by_1,\dots,\by_{c-1})} \ ,
\qquad
\Z(\by_1,\dots,\by_{c-1}) = 
\sum_{\by} w_{\rm link}(\by) w_{\rm iter}(\by,\by_1,\dots,\by_{c-1}) \ .
\label{eq_PZ}
\eeq
Constructed in this way $P(\by|\by_1,\dots,\by_{c-1} )$ is a conditional 
probability distribution over $\by$. Given the values of the hopping 
trajectories $\by_1,\dots,\by_{c-1}$ it is actually possible to perform an
exact sampling from $P(\by|\by_1,\dots,\by_{c-1} )$ and to compute the
normalization constant $\Z(\by_1,\dots,\by_{c-1})$. In fact this reduces to
a relatively simple single site problem: one has to construct the occupation 
trajectory $n(\t)$ of a site, and the associated hopping trajectory $\by$
towards one of its neighbors, given the hopping trajectories
$\by_1,\dots,\by_{c-1}$ on the other adjacent edges (see Fig.~\ref{fig_traj_iter}
and recall the pictorial
representation given for the Ising case in Fig.~\ref{fig_Ising_iter}).
The $c-1$ hopping trajectories on the upper edges impose
that $n(\t)$ changes by $\pm 1$ at the times particles arrive or depart
from the considered central site. Between these times we shall show
in Sec.~\ref{sec_derivation} that the single site problem is described by an
effective Hamiltonian $V(n)-J(a+a^\dag)$; each change in the
value of $n(\t)$ provoked by the creation/annihilation operators in this
effective Hamiltonian is associated to an hopping event in the new (downwards)
trajectory $\by$. An example of this construction is given in
Fig.~\ref{fig_traj_iter}. 
The computation of $\Z(\by_1,\dots,\by_{c-1})$ can be performed recognizing it
as the partition function of the single-site effective Hamiltonian.

\medskip

Suppose now a representation of
$\eta_{\rm cav}$ is available under the form (\ref{eq_rep_popu}); one can plug
it into the right-hand-side of Eq.~(\ref{eq_fpoint_number2}) and construct
a new set of trajectories and weights corresponding to its left-hand-side, let us call
them $\{\by'_j, g'_j\}_{j=1}^\Nt$. Independently and
identically for each $j \in [1,\Nt]$, this can be done by:
\begin{itemize}
\item drawing $c-1$ integers $i_1,\dots,i_{c-1} \in [1,\Nt]$, each of them 
independently with probability $g_i$
\item drawing $\by$ from $P(\by|\by_{i_1},\dots,\by_{i_{c-1}} )$
\item setting $\by'_j = \by$, $g'_j = \Z(\by_{i_1},\dots,\by_{i_{c-1}})$
\end{itemize}
The new weights $g'_j$ are then normalized to fulfill Eq.~(\ref{eq_norm}).
This process can be 
iterated, plugging in the r.h.s. of Eq.~(\ref{eq_fpoint_number2}) the newly
obtained representation $\{\by'_j, g'_j\}$, and so on and so forth. After a 
certain number of steps, starting from an arbitrary initial 
condition\footnote{The initial condition should however allow for the
  possible spontaneous symmetry breaking $\la a \ra \neq 0$; we shall discuss
  this point further in Sec.~\ref{sec_derivation}.}, a stationary
solution is reached. 

The physical observables can be computed from this representation of 
$\eta_{\rm cav}$. Let us first explain the determination of the
average occupation number of one site. Following our convention
we denote $\bn$ the occupation number imaginary-time trajectory $n(\t)$.
Its probability distribution $\eta(\bn)$ is expressed by considering the
complete environment of a vertex with its $c$ neighbors, as was done for the
local magnetization of the Ising model in Eq.~(\ref{eq_Ising_final}) and
schematized in Fig.~\ref{fig_Ising_final}, 
\beq
\eta(\bn) = \frac{1}{z_{\rm site}} \sum_{\by_1,\dots,\by_c} 
\eta_{\rm cav}(\by_1) \dots \eta_{\rm cav}(\by_c) \,
w_{\rm site}(\bn,\by_1,\dots,\by_c) \ .
\label{eq_number_final}
\eeq
The explicit expression of the weight $w_{\rm site}$ shall be given in
Sec.~\ref{sec_derivation}. It is however intuitive that it will contain a
factor $\exp[-\int \dd \t V(n(\t))]$ arising from the local part of the
Hamiltonian, and requires a consistency between $\bn$ and
$(\by_1,\dots,\by_c)$. In fact all the discontinuities in the occupation
trajectory $n(\t)$ are fixed by the hopping events on the $c$ neighboring
edges. If the number of hoppings towards the considered site equals the number
of jumps outside, then the trajectory $n(\t)$ is fixed up to a global shift
$n(\t)\to n(\t) + m$ with $m$ independent of time. Whenever there is
an unbalanced number of hopping towards/outside the vertex no periodic
trajectory $n(\t)$ can be constructed, $w_{\rm site}=0$ for such a
configuration of $(\by_1,\dots,\by_c)$. The average occupation number of a site
is easily obtained from this probability $\eta(\bn)$,
\beq
\la a^\dag a \ra = 
\sum_{\bn} \eta(\bn) n(0)
= \frac{\underset{\by_1,\dots,\by_c}{\sum} 
\eta_{\rm cav}(\by_1) \dots \eta_{\rm cav}(\by_c) 
\underset{\bn}{\sum} w_{\rm site}(\bn,\by_1,\dots,\by_c) n(0)}
{\underset{\by_1,\dots,\by_c}{\sum} 
\eta_{\rm cav}(\by_1) \dots \eta_{\rm cav}(\by_c) 
\underset{\bn}{\sum} w_{\rm site}(\bn,\by_1,\dots,\by_c) } \ ,
\label{eq_nmed}
\eeq
where in the last expression we have explicited the normalization constant
$z_{\rm site}$. Note that the time $\t=0$ where $n(\t)$ is evaluated is
arbitrary because of the cyclic invariance along the imaginary time axis,
hence $n(0)$ can be equivalently replaced by the temporal average
$\frac{1}{\beta} \int_0^\beta \dd \t \, n(\t)$.
Given a representation of $\eta_{\rm cav}$ as a weighted sample
(\ref{eq_rep_popu}) the numerical determination of $\la a^\dag a \ra$ is
straightforward. Both the numerator and the denominator of (\ref{eq_nmed})
have the form of the average of a function $u(\by_1,\dots,\by_c)$ with
the $\by$'s independently drawn from their distribution $\eta_{\rm cav}$. As
already explained drawing from $\eta_{\rm cav}$ a trajectory $\by$ amounts to
draw an index $i\in[1,\Nt]$ with probability $g_i$ and picking the element
$\by_i$ of the population. In formula this method of sampling leads to
\beq
\sum_{\by_1,\dots,\by_c} \eta_{\rm cav}(\by_1) \dots \eta_{\rm cav}(\by_c)
\, u(\by_1,\dots,\by_c) = \frac{1}{\Ntry} \sum_{j=1}^\Ntry
u(\by_{i_{j,1}},\dots,\by_{i_{j,c}}) \ ,
\label{eq_avg_g}
\eeq
where $\{i_{j,l}\}$ are integers drawn independently with the probability
$g_i$, with $j\in[1,\Ntry]$ and $l\in[1,c]$. The numerical accuracy of the
sampling is increased by taking a large value of $\Ntry$. Moreover one can
interleave average steps with iteration steps, determining a new set of
weights $g_i$ and trajectories $\by_i$ and recomputing independent averages
on this new representation of $\eta_{\rm cav}$.

We shall show in Sec.~\ref{sec_derivation} that all the other observables
defined in Sec.~\ref{sec_definition} (order parameter $\la a \ra$,
free-energy, kinetic/potential energy and Green function) can also be
expressed as averages of the form (\ref{eq_avg_g}), and thus can be
efficiently computed by the method developed here.

\subsection{Results}
\label{sec_results}

\begin{figure}
\includegraphics[width=8cm]{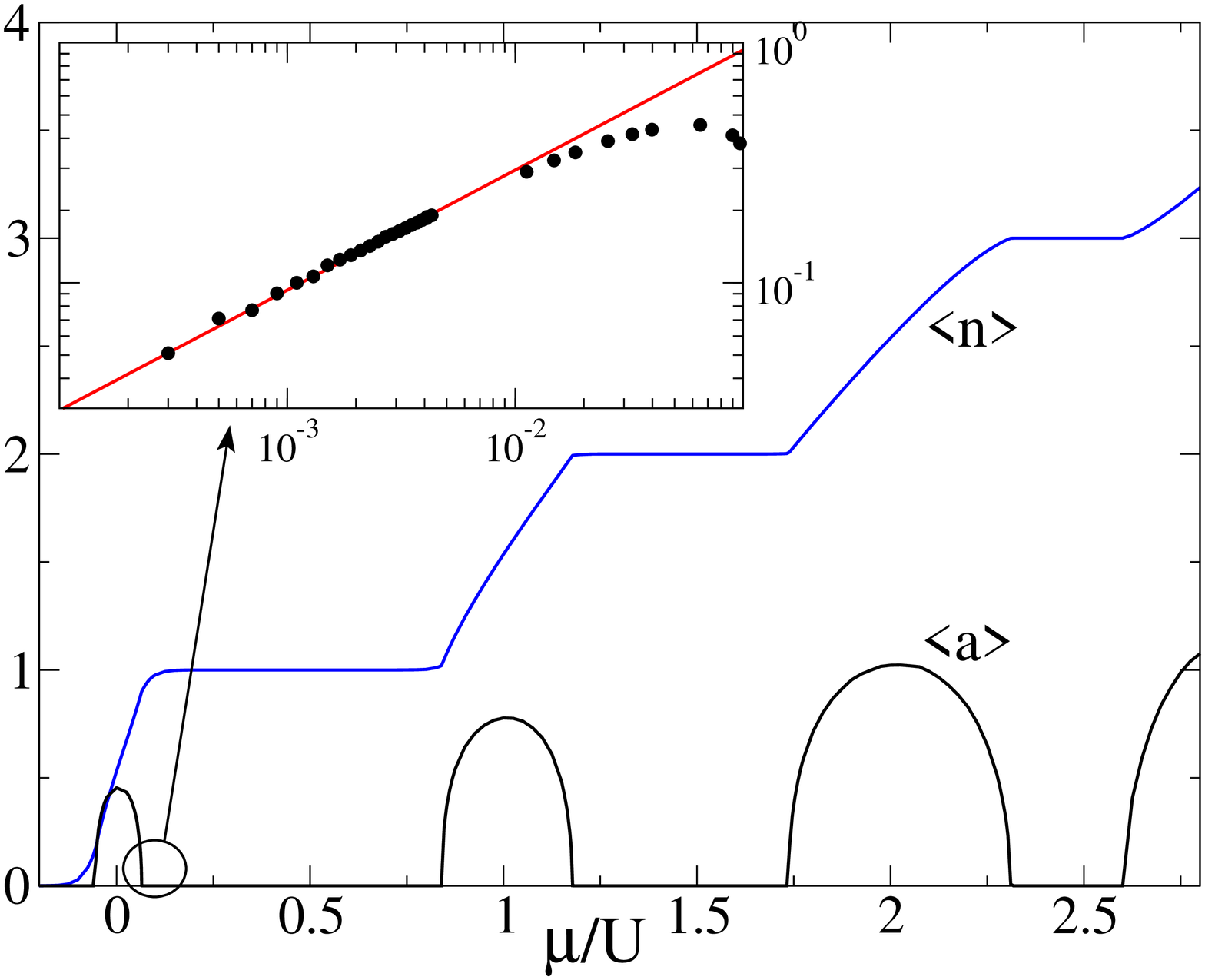}
\caption{{\it (Main panel)} $\la n \ra$ and $\la a \ra$ as functions 
of $\mu/U$ for $J/U=0.02$, at $\b J =1$ and $N_{\rm cut}=6$ for Bethe lattices
of connectivity $c=4$.
Full lines are obtained by joining points measured every $\D \mu/U \sim 0.05$, except close
to the critical points where $\D \mu/U \sim 0.001$.
{\it (Inset)} Blow-up of the region close to the first SF-MI transition that happens at $\m_c = 0.0648 \, U$.
Here $\la a \ra$ is plotted as a function of $\d = (\mu_c - \mu)/U$ (circles). The full line is a fit to
$\la a \ra \sim A \d^{0.5}$ with $A = 2.9475$.}
\label{fig_n_a}
\end{figure}

In the following we present the results obtained for the Bose-Hubbard model
on the Bethe lattice
solved with the cavity method. 
We have shown in the previous sections that 
{\it in the thermodynamic limit}
the exact resolution of the model amounts to solve
Eq.~(\ref{eq_fpoint_number}) and we explained how this can be done
with arbitrary numerical precision
using the population dynamics algorithm described in 
subsection~\ref{sec_qcav_number_eq},
encoding $\eta_{\rm cav}(\by)$
as a population of a large number $\Nt$ of trajectories. 
We typically used $\Nt=32768$ and checked that the results 
were unchanged when using larger $\Nt$.
We recall that in the thermodynamic limit the local observables
(say, $\la n_i \ra$) are independent on the site $i$.
In the following we focus on Bethe lattices of connectivity 
$c=4$ and $6$, which mimic the connectivity
of the two and three-dimensional square and cubic lattices, respectively.

Note that in order to implement the summations over $n$ involved in the
computation of the various weights appearing in Eqs.~(\ref{eq_fpoint_number})
and (\ref{eq_number_final}), we have to introduce a cutoff on the possible values
of $n$, $0\leq n < N_{\rm cut}$. The technical details will be discussed in
section~\ref{sec_derivation}.
Since all the results presented below corresponds to regions of the parameter
space such that $0 \le \langle n \rangle \le 3$, we work with a cut-off on the 
occupation number in the interval $4 \leq N_{\rm cut} \leq 6$. 
We have checked for various values of the parameters
that all the results are stable with respect to changes of the cutoff (in some
cases also up to $N_{\rm cut} > 6$).

In this section we also compare our results with other analytic
approaches, such as the variational mean-field treatment described
in Sec.~\ref{sec_rev_MF} 
and the bosonic Dynamical Mean Field Theory (B-DMFT)\footnote{
Note that we compare our results for a Bethe lattice with finite 
connectivity with those obtained in~\cite{BDMFT_2} using the
semi-circular density of states. The latter is strictly valid in the limit
of large connectivity, but this effect should be of subleading order in the
large connectivity expansion.} 
discussed in Sec.~\ref{sec_BDMFT}, that correspond to the large $c$ 
limit of our method~\cite{BDMFT_1,BDMFT_2}.
We also compare them to Quantum Monte Carlo simulations on $2d$ 
square lattice~\cite{bh_qmc_2d} and
$3d$ cubic lattice~\cite{bh_qmc_3d}.

\subsubsection{Thermodynamic observables}

Once the numerical iterative procedure has converged to a fixed point solution,
one can compute the thermodynamic observables.
Fig.~\ref{fig_n_a} shows the behavior of $\langle n \rangle$ and
$\langle a \rangle$, as a function of the chemical potential, $\mu / U$
for $J/U=0.02$, and $\beta J=1$ for a Bethe lattice of connectivity $c=4$. 
We have also computed $\la n \ra$ and $\la a \ra$
at lower values of the temperature ($\beta J=2$, $4$, and $6$), but we find
that the results are unchanged within our numerical accuracy, except very close
to the lowest-$\mu$ superfluid-insulator transitions, where a little temperature
effect is detected\footnote{
In particular the cusp in the
kinetic energy at transition
from the SF to the MI becomes sharper and $e_K$
becomes slightly smaller as we go from $\beta J=1$ to $\beta J=2$ in 
the first superfluid region.
No further modifications are observed within our numerical errors 
for $\beta J > 2$.
}
between $\b J =1$ and $\b J=2$ (while no change is detectable above
$\b J=2$). Hence, we can safely assume that for $\b J \geq 2$ we are in the zero
temperature regime.

The plot of Fig.~\ref{fig_n_a} clearly shows a sequence of transitions from
superfluid phases (SF) to Mott insulators (MI).
The superfluid phase is characterized by a finite value of $\langle a \rangle$,
which corresponds to a finite condensate fraction $f_{\rm BEC} = 
\langle a \rangle^2/\langle n \rangle$, while
in the Mott insulator $\langle a \rangle = 0$. 
In the limit of zero temperature in the MI regions 
the average number of bosons per site, $\langle n \rangle$
is fixed to integer 
values, $\langle n \rangle = 1$, $2$, $3$, $\ldots$.
As a result, the compressibility of the system, $\chi 
= \partial \langle n \rangle
/ \partial \mu$ vanishes identically.
At small but finite temperature, we expect the
compressibility to be exponentially small in the temperature, and indeed
in Fig.~\ref{fig_n_a} we observe that the compressibility at commensurate
density is practically zero even at $\b J=1$.

As shown in the inset of Fig.~\ref{fig_n_a}, the critical exponent for 
$\la a \ra \sim (\mu_c - \mu)^{\epsilon}$ at the transition from the SF phase 
to the MI is consistent with the mean-field value $\epsilon = 1/2$.
As in the classical case, the critical exponents found on the Bethe lattice are
equal to the mean-field ones (see~\cite{qcav_our} for a detailed discussion
in the case of the quantum spin-1/2 ferromagnet on the Bethe lattice). 
Therefore, we also expect the other critical 
exponents to coincide with their mean-field values.

\begin{figure}
\includegraphics[width=8cm]{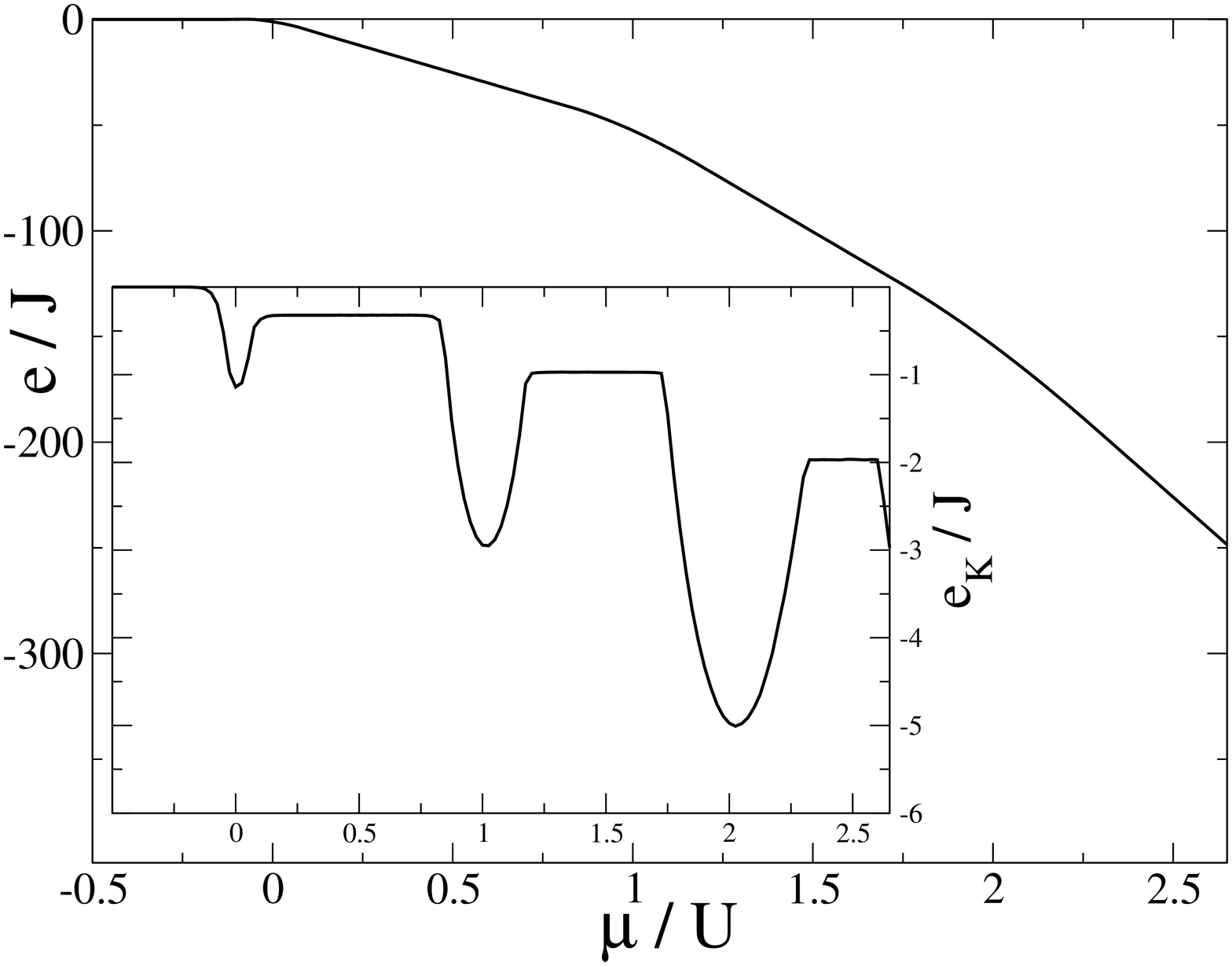}
\caption{{\it (Main panel)} Total energy, $e = \langle H \rangle / N$, as a function 
of $\mu/U$ for $J/U=0.02$, at $\b J =1$ and $N_{\rm cut}=6$ for Bethe lattices
of connectivity $c=4$.
{\it (Inset)} Kinetic energy $e_K$ as a function of
the chemical potential for the same values of the parameters.}
\label{fig:energy}
\end{figure}

In Fig.~\ref{fig:energy} we plot the total energy,  
$e = \langle H \rangle/N$, 
as a function of the chemical potential 
for the same values of the parameters as
in Fig.~\ref{fig_n_a}. 
We have also computed the free energy of the system, $f$, but
we do not show it since 
at these low temperatures the difference between the energy and the free 
energy is very small, and practically
undetectable within our numerical precision, which confirms that we
are effectively in the zero-temperature regime and the system can be considered
to be in its ground state.
In the inset of 
Fig.~\ref{fig:energy} we show the behavior of the kinetic energy,
$e_K$ as a function of $\mu/U$.
The kinetic energy is, of course, lower in the SF phase than 
in the MI.
Moreover we find that $e_K$ does not vanish in the MI phase,
differently from the traditional mean-field approach, where hopping is
completely suppressed in this phase.
Note that both the kinetic energy and the potential energy
show a discontinuity in the first derivative at the transition between
the MI and the SF phase.
On the contrary, the first derivative of the total energy (and of the
free energy as well) is continuous at the transition;
only its second derivative shows a discontinuity.

\subsubsection{Phase diagram}

\begin{figure}
\includegraphics[width=8.5cm]{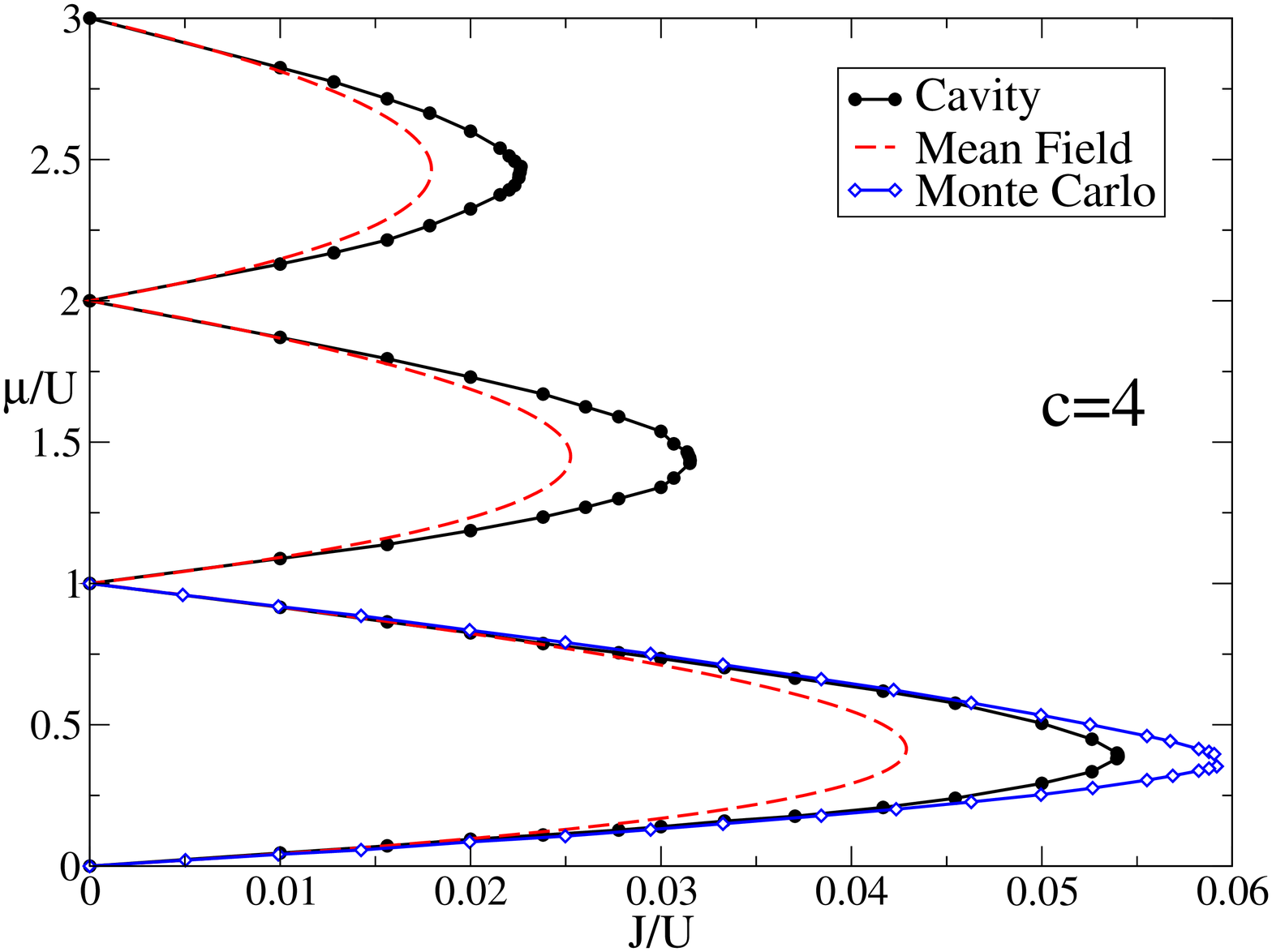}
\includegraphics[width=8.5cm]{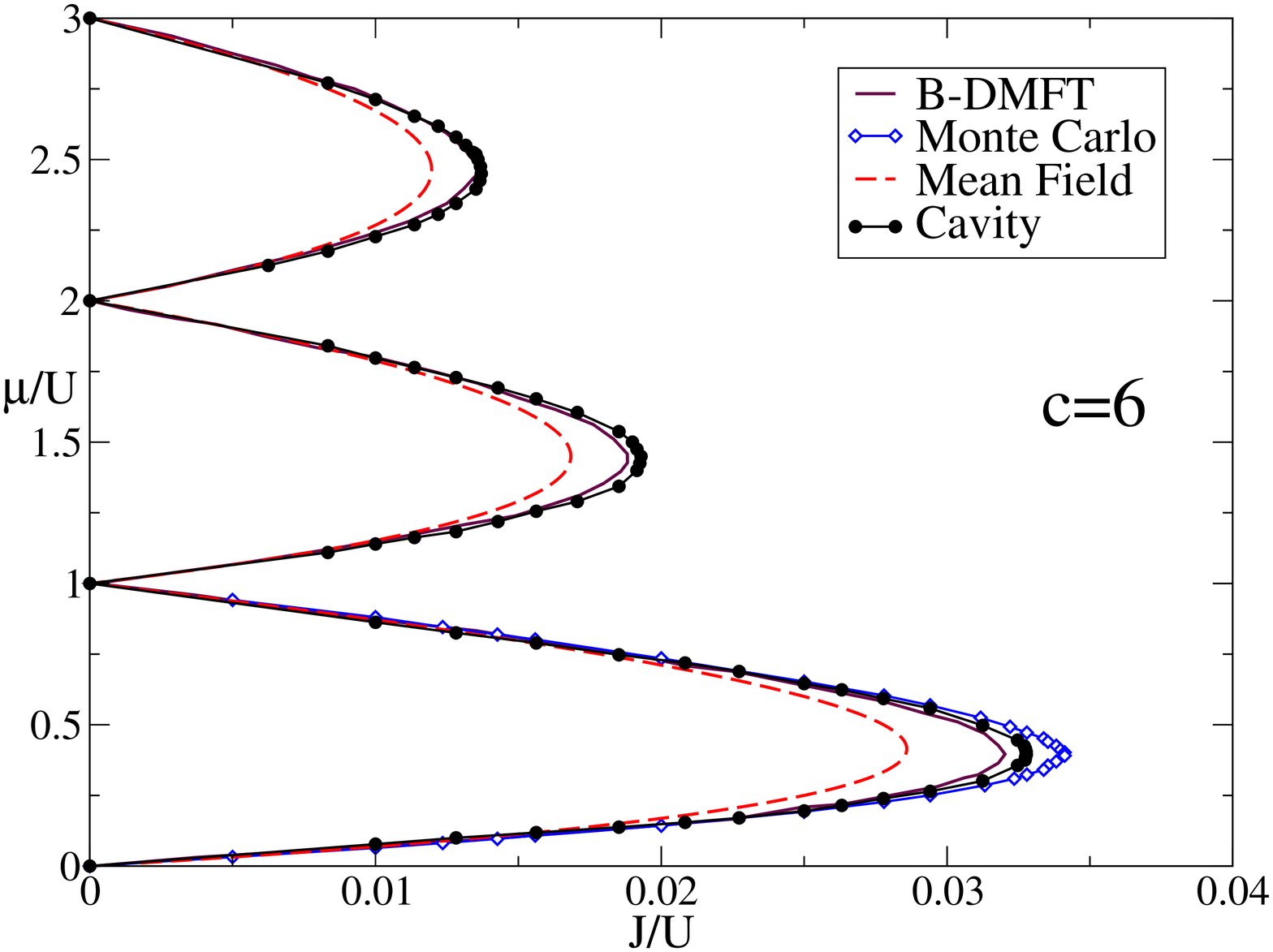}
\caption{
``Zero temperature'' phase diagram of the Bose-Hubbard model on the Bethe lattice
in the $(J/U,\mu/U)$ plane at connectivity $c=4$ {\it (left panel)} and 
$c=6$ {\it (right panel)}. The results obtained within the cavity method 
(black lines and filled circles)
are compared to Monte Carlo simulation (blue lines and open diamonds) 
on the square (data from~\cite{bh_qmc_2d}) and the cubic 
(data from~\cite{bh_qmc_3d}) lattice, and to the prediction of the
mean-field approach (red dashed curves) 
and of bosonic DMFT  (brown full curve in right panel, 
data from~\cite{BDMFT_2}).}
\label{fig:Mott-pdz4_6}
\end{figure}

In Fig.~\ref{fig:Mott-pdz4_6} we present the ``zero temperature'' 
(recall that our method only works at finite temperature, hence we use
quotes to remind that an extrapolation to $T=0$ is needed)
phase diagram of the Bose-Hubbard model on the Bethe lattice
of connectivity $c=4$ (left panel) and $c=6$ (right panel) in the
$(J/U,\mu/U)$ plane. The phase boundaries have been obtained by computing
$\la a \ra$ as a function of $\mu/U$ at constant $J/U$. 
In this case we used $4 \leq N_{\rm cut} \leq 6$ and $4 \leq \beta J \leq 6$
(lower values of $\b J$ and higher values of $N_{\rm cut}$ 
have been used at higher $\mu$). As already
discussed, at these values of $\beta$ we are effectively in the
zero temperature regime.
The lobelike shape of the phase boundary between MI and SF phases 
is qualitatively similar to the one found in the mean field approach.
In Fig.~\ref{fig:Mott-pdz4_6} we also compare the results found with the 
quantum cavity method with the outcomes of Quantum Monte Carlo simulations 
(on the square lattice
for $c=4$~\cite{bh_qmc_2d} and on the cubic lattice for $c=6$~\cite{bh_qmc_3d}),
and with the results
of other analytical approaches. This comparison shows that the cavity
method does a fairly good job in locating the lobelike 
contours between the MI and the SF phase and performs much better than 
the mean-field approach.
It also performs slightly better than the Bosonic DMFT~\cite{BDMFT_2}
for $c=6$,
although the difference
between the cavity method and the B-DMFT is expected 
to become small 
as the connectivity is increased. It would be interesting in this respect to
compare the cavity method and B-DMFT for $c=4$.

\begin{figure}
\includegraphics[width=8.5cm]{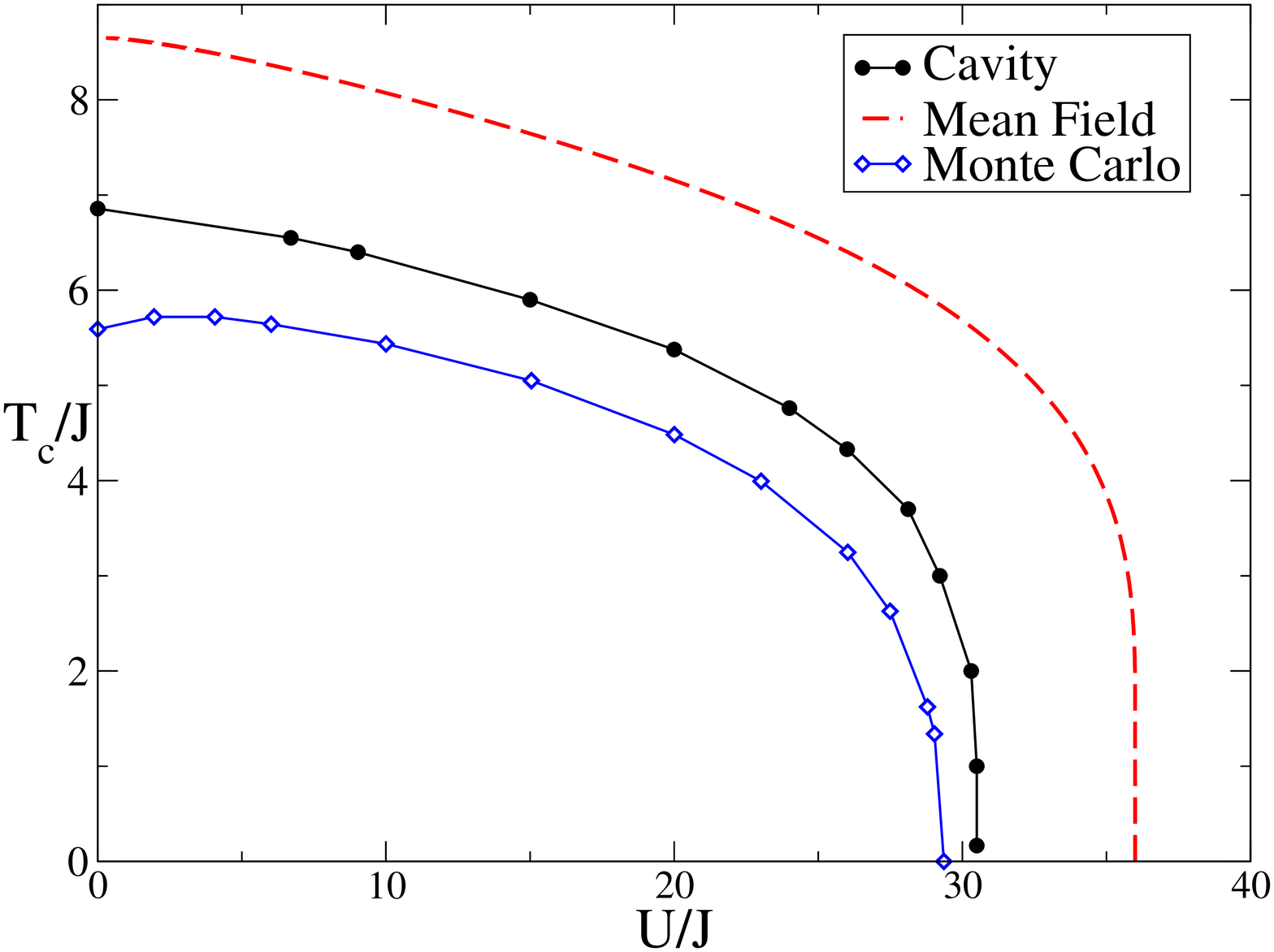}
\caption{Interaction dependence of the transition temperature at unit
filling. The result obtained on the $c=6$ Bethe lattice 
(black lines and filled circles)
is compared to Monte Carlo simulation (blue lines and open diamonds) 
on the cubic lattice (data from~\cite{bh_qmc_3d}), and to the prediction of the
mean-field approximation (red dashed curves).}
\label{fig_PD_Tpos}
\end{figure}

In addition to the low temperature phase diagram we have computed the 
transition temperature line from the ``insulating'' phase 
(which we defined by $\la a \ra =0$, but has still a finite conductivity at finite temperature)
to the superfluid one (defined by $\la a \ra \neq 0$)
at unit filling ($\la a^\dag a \ra =1$), for Bethe lattices of
connectivity $c=6$. This result is plotted in
Fig.~\ref{fig_PD_Tpos} and compared to the mean-field prediction and
to the three-dimensional Monte-Carlo simulations of~\cite{bh_qmc_3d}.

\subsubsection{Green functions and particle occupation statistics}

\begin{figure}
\includegraphics[width=8.5cm]{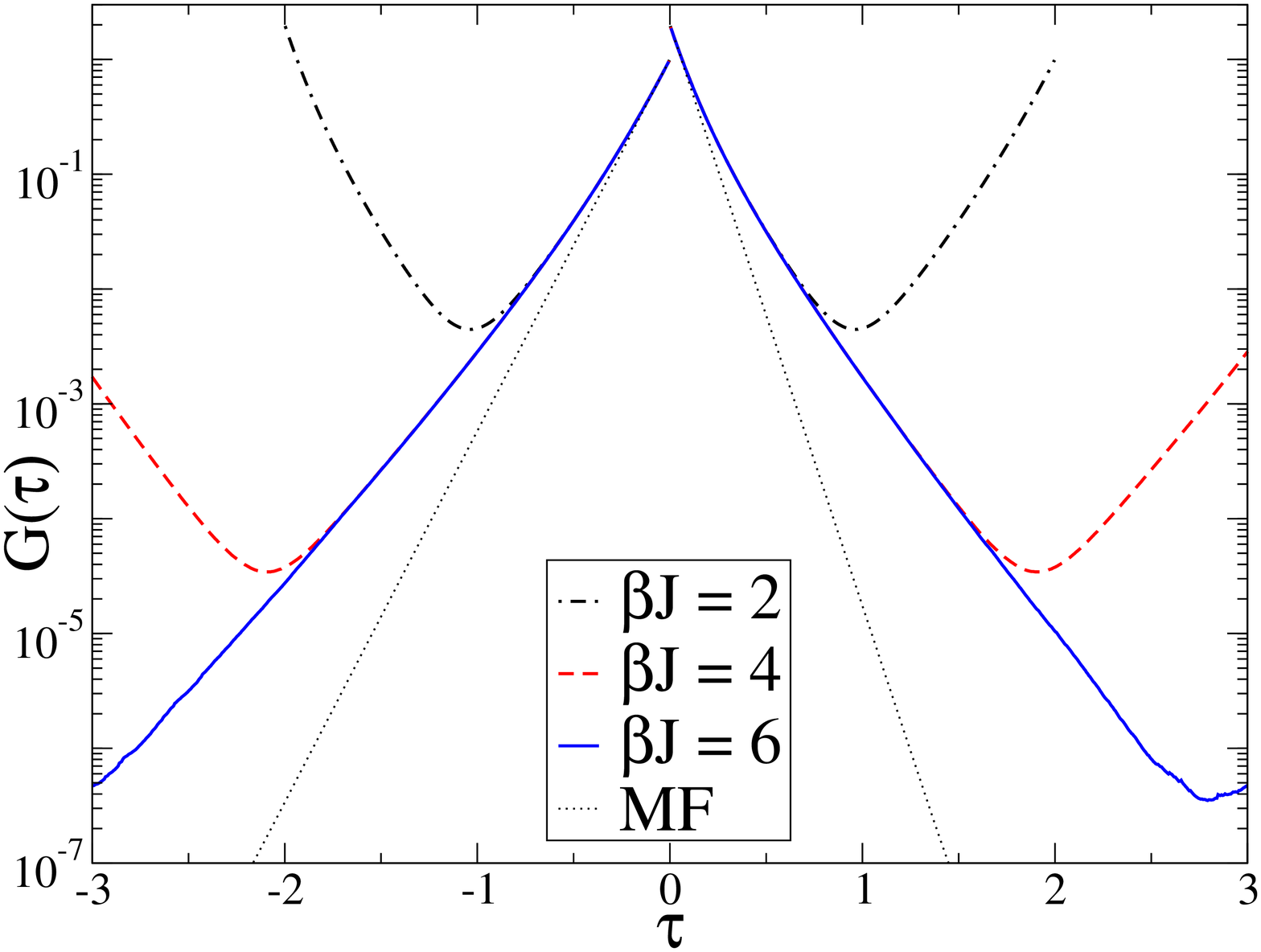}
\includegraphics[width=8.5cm]{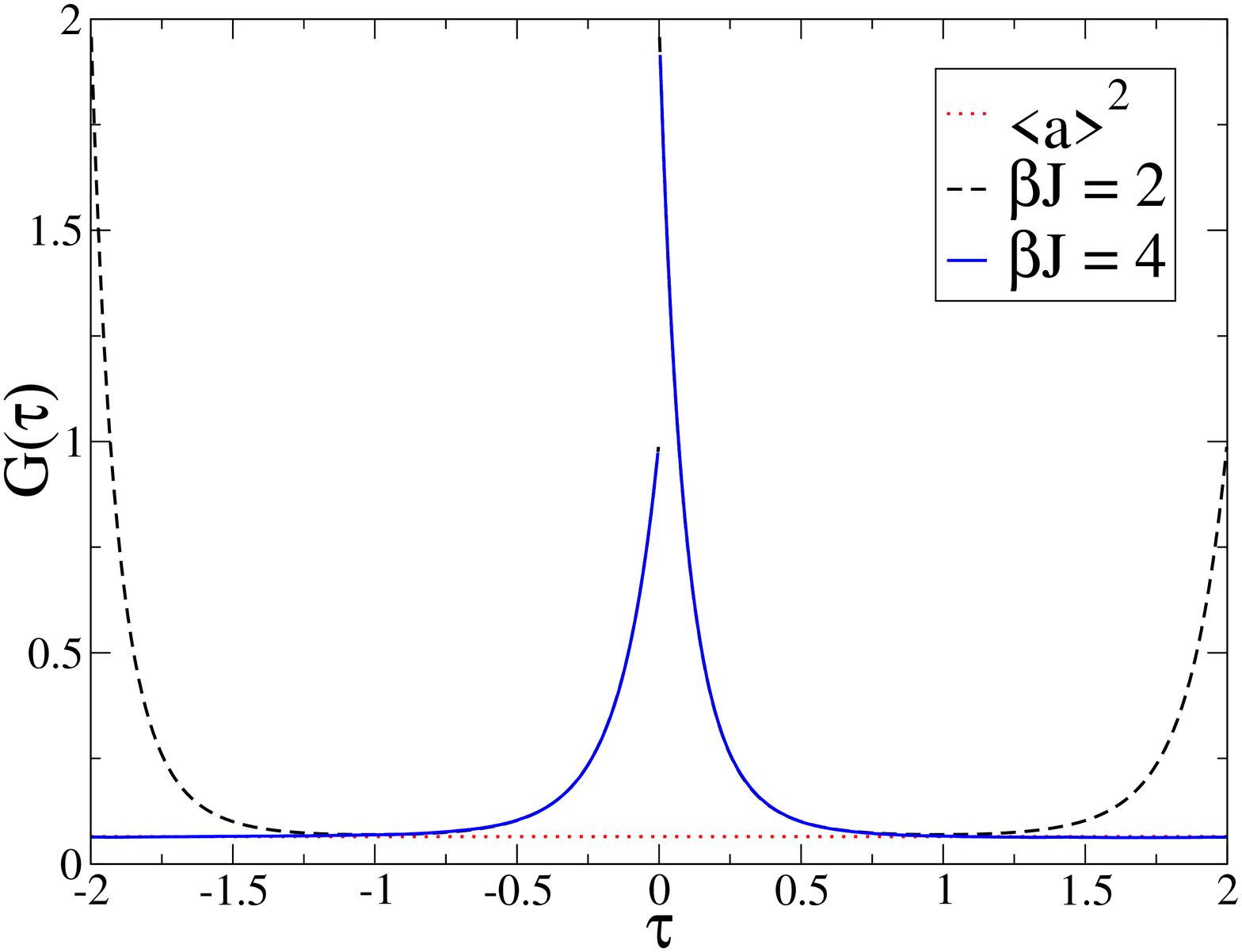}
\caption{Green function $G(\t)$ as a function of the 
imaginary time at $c=4$.
{\it (Left panel)} 
Results in the MI phase, at $J/U=0.0523$, $\mu/U=0.39$, $N_{\rm cut}=4$ 
and $\beta J=2$ (black dashed-dotted curve), 
$4$ (red dashed curve), and $6$ (blue continuous curve). 
Also reported is the mean-field result for the same
parameters (black dotted curve).
{\it (Right panel)} 
Results in the SF phase, at $J/U=0.0555$, $\mu/U=0.39$, $N_{\rm cut}=6$ 
and $\beta J=2$ (black dashed curve) and $4$ (blue curve). Also reported is the long
time limit of $G(\t)$, which corresponds to the value 
of $\la a \ra^2$ for these values of the parameters.
}
\label{fig:Green}
\end{figure}

As far as the results shown up to now are concerned, 
there are no qualitative differences between the Bethe lattice
and the fully connected models (or equivalently the mean field 
approximation). We shall now present the results for
the one-particle on-site imaginary time Green function, which exhibits
a richer behavior with respect to the mean-field description.
This quantity, defined in Eq.~(\ref{eq_green_def}) as
$G(\t)=\langle T a(\t) a^{\dagger}(0) \rangle$, is plotted in 
Fig.~\ref{fig:Green} for two set of values of the parameters $(J/U,\mu/U)$, 
one in the MI phase, the other in the SF, both very close to the tip of the 
first insulating lobe.

Let us first remark that the decay of the Green function is independent
on temperature in this temperature range, the effect of temperature
being only to cut the decay around $|\t| \sim \beta/2$ due to
periodicity. Therefore we can safely assume that the Green functions
reported in Fig.~\ref{fig:Green} are good estimates of their 
zero-temperature limit.

For the interpretation of these curves it is worth recalling the
spectral representation of the Green function~\cite{Negele} at zero
temperature:
for $\t > 0$, $G(\t) = |\la a \ra|^2 + \int_0^\io \r_p(E) \exp(-E \t)$,
where $\r_p(E)$ is the (on-site) spectral density for particle excitations.
A similar expression holds for $\t<0$ in terms of hole excitations.
The long time limit of $G(\t)$ is thus strictly positive in the SF phase,
and we checked that is in agreement with the value $|\la a \ra|^2$ previously
computed (see the dotted line in the right panel of Fig.~\ref{fig:Green}).

Let us now turn to the discussion of the Green function in the MI.
In the mean-field description of this phase the hopping is completely
suppressed, hence particle (resp. hole) excitations corresponds to the
addition (resp. removal) of one particle from the background of the
commensurate filling of immobile particles. To be concrete let us
consider the first lobe with $\langle n \rangle =1$; the energy of these 
particle and hole excitations are easily found to be $E_p=U-\m$ and 
$E_h=\mu$, hence follows the expression of the Green 
function~\cite{fc_Green}:
\beq
G_{\rm mf}(\t) = \theta(\t) 2 e^{-(U-\m)\t} + \theta(-\t) e^{-\m \t} \ .
\eeq
In other words the spectral density of particle excitations is made of
a single delta function (and similarly for the hole excitations).
The mean-field Green function is reported in Fig.~\ref{fig:Green} as a 
dashed black line; the comparison with the results of the Bethe lattice
computations shows that the latter is more complex. The decay of the latter 
is not a simple exponential, thus reflecting a non-trivial spectrum of 
excitations. Moreover the energy of particle and hole excitations is lower on 
the Bethe lattice, as can be seen from their slower decay.
This happens because on the Bethe lattice, 
due to the local finite connectivity,
particles can still hop around even in the MI phase, resulting in a gain
of kinetic energy and a lowering of the energy cost of the excitations.
Note that different improvements of the mean-field treatment, for instance
Random Phase Approximations (RPA)~\cite{Sengupta05,Menotti08},
provide a much better description of the spectral density in the Mott phase, in particular a good approximation to the wave-vector dependence of
the excitation energies for finite-dimensional lattices. A quantitative comparison with the predictions of the RPA is beyond the scope of this paper. We focused instead on the comparison with the fully-connected result and B-DMFT, because they are related to the Bethe lattice via a well-controlled limit, namely the large connectivity one.

We did not attempt to perform the inverse Laplace transform
to determine the spectral function, yet we performed the following analysis
to estimate the scale $E_p$ of the slowest decay in the $\t >0$ part of the
Green function. A plot of $\log[ \t^\a G(\t) ]/\t$ is seen to
approach a constant at large $\t$ for $\a \sim 1$, suggesting that
$G(\t)$ decays as $\exp(-E_p \t)/\t$, hence that the spectral function
$\r_p(E)$ is finite at $E=E_p$ and vanishes for $E<E_p$. This is consistent with
the results of~\cite{Sengupta05,Menotti08}.
For simplicity, we have chosen to fit
the Green function to $G(\t) = 2 e^{-E_p \t} (1- e^{-\D_p \, \t})/(\D_p \, \t)$, 
corresponding
to a flat density of states in $[E_p,E_p + \D_p]$. We stress that this is
an arbitrary choice that we made only to fit the long time behavior of
$G(\t)$ and determine $E_p$. Note that at very small $\t \ll 1/J$, hopping
is irrelevant and $G(\t) \sim G_{\rm mf}(\t)$ at first order in $\t$. 
This imposes a relation between $\D_p$ and $E_p$, hence there is a single 
free parameter in our fitting procedure.

\begin{figure}[t]
\includegraphics[width=8.5cm]{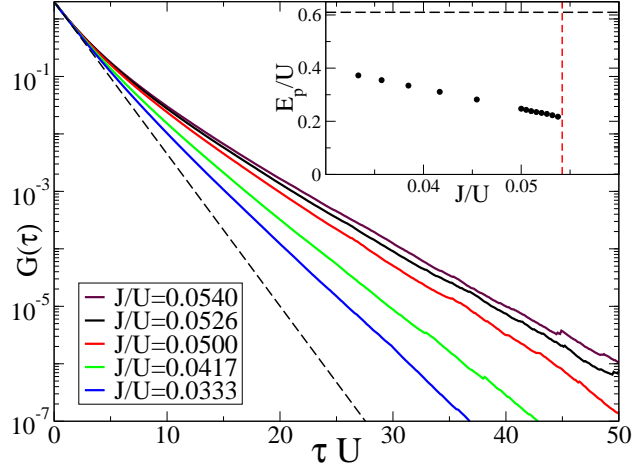}
\caption{
{\it (Main Panel)}
Green function $G(\t)$ as a function of the rescaled
imaginary time $\t U$ for $c=4$ in the MI phase, close to the tip
of the first lobe at $\mu/U = 0.39$, $N_{\rm cut}=4$ 
and $\beta J=6$, at different values of $J/U$
(approaching the transition to the superfluid, from bottom
to top).
Also reported is the mean-field result for the same
parameters (black dashed curve).
{\it (Inset)} Particle excitation gap $E_p/U$ is plotted
as a function of $J/U$ (see text for details); the dashed black
line is the mean field result $E_p/U = 1-\m/U$, 
the vertical line marks the transition
to the superfluid.
}
\label{fig:GreenMI}
\end{figure}

In Fig.~\ref{fig:GreenMI} we show the behavior of the 
Green function on approaching the phase transition from
the MI phase, close to the tip of the first lobe by varying
$J/U$ at constant $\m/U = 0.39$.
We focus on positive $\t$ and fix $\b J = 6$, which according
to Fig.~\ref{fig:Green} is a low enough temperature, such that
the Green function is identical to its zero-temperature limit for
$\t < \b/2$. We use the rescaled time variable $\t U$, in such a way
that the mean field result $G_{\rm mf}(\t) = 2 \exp[-\t U (1-\mu/U)]$ yields
the same decay at all values of $J/U$. On the contrary,
the Bethe lattice Green function depends quite strongly on
$J/U$, and its decay slows down on approaching the transition. The result
of the fitting procedure explained above is shown in the inset of 
Fig.~\ref{fig:GreenMI} and shows that the parameter $E_p$
is smaller by a factor of 2 on the Bethe lattice compared to the mean field
results; it decreases on approaching the transition, but remains finite 
at the transition. Indeed we have computed here the on-site Green-function,
which reflects only the localized single-particle excitations, whereas
the transition towards the superfluid phase is towards a delocalized state
(small momentum in the finite dimensional case~\cite{bh_qmc_3d}). To detect 
the signature of the transition in such a way it
would be necessary to compute the spatial and temporal dependence of the
Green function, which is in principle possible using the cavity method (see next subsection).

We expect (but did not check in detail) that also
in the SF phase the on-site Green function will decay exponentially
up to the transition point. This shows also that the BEC state on the
Bethe lattice has a peculiar nature as compared to the finite dimensional
case: condensation happens in the uniform state, which is separated from the
other eigenvalues of the connectivity matrix by a gap~\cite{gap1,gap2}. In particular,
one can check explicitly by a Bogoliubov-like computation~\cite{AGDbook} that the
single-particle excitation spectrum is gapped at small $U/J$.

The difference between the mean field approach and
the Bethe lattice result is also unveiled by the analysis of the
probability distribution $P_n$, already studied recently 
in~\cite{bh_qmc_number} using
Monte Carlo simulations, defined 
as the probability to detect $n$ bosons on a given lattice site:
\beq
P_{n_0} = \la \delta_{n, n_0} \ra = 
\frac1Z \Tr \left[ \delta_{n,n_0} e^{-\b H} \right] \  .
\eeq
This is another quantity sensitive to the occupation
number fluctuations. In table~\ref{tab_Pn} 
we report the values of $P_n$ in the MI and in the SF, for the same
values of the parameters as for Fig.~\ref{fig:Green}.
\begin{table*}[t]
\begin{tabular}{|c|c|c|c|c|c|c|}
\hline
$n$         & 0 & 1 & 2 & 3 & 4 & 5      \\
\hline
$P_n$ (SF)  & 0.034  & 0.93  & 0.034  & 0.00027  & 5.7e-07  &   2.2e-11             \\
$P_n$ (MI)  & 0.023  & 0.95  & 0.023  & 0.00013  &   &     \\
\hline
\end{tabular}
\caption{
Values of $P_n$ in the MI and in the SF, for the same
values of the parameters as in Fig.~\ref{fig:Green}, 
namely $J/U=0.0523$, $\mu/U=0.39$, $N_{\rm cut}=4$ 
and $\beta J=4$ (in the MI phase)
and $J/U=0.0555$, $\mu/U=0.39$, $N_{\rm cut}=6$ 
and $\beta J=4$ (in the SF phase).
}
\label{tab_Pn}
\end{table*}
These data clearly indicate that even in the MI phase the occupation 
number of 
a given lattice site is not strictly fixed to an integer value, as in the
mean field approach: there is still a finite probability of finding a number
of bosons different from one within the lobe at $\la n \ra = 1$.
Interestingly enough, these quantities can be probed in experiments~\cite{multi_occ}.

\subsubsection{One particle density matrix}

\begin{figure}[t]
\includegraphics[width=8.5cm]{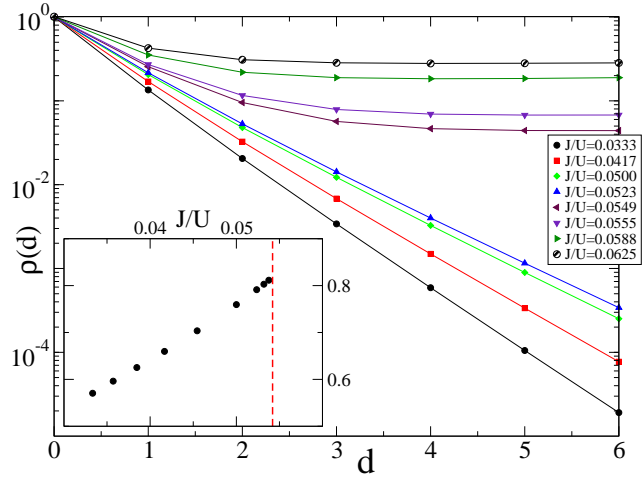}
\caption{
{\it (Main Panel)} 
One particle density matrix $\r_d = \la a^\dag_0 a_d \ra$ at $c=4$, fixed
$\mu/U=0.39$, $\beta J=2$, $N_{cut}=4$, and different values of
$J/U$ crossing the tip of the first MI lobe in Fig.~\ref{fig:Mott-pdz4_6}.
In the MI phase $\r_d$ decays to zero, while in the SF phase it decays
to $|\la a \ra|^2$.
{\it (Inset)} Correlation length as a function of $J/U$ in
the MI phase. The vertical line marks the MI-SF transition.
}
\label{fig:DM}
\end{figure}

We present now our results for the one-particle density matrix
$\r_{ij} = \la a^\dag_i a_j \ra$. 
Because of the invariance structure of the Bethe lattice
this will depend (in the thermodynamic limit) only on the distance between
the sites $i$ and $j$, that we denoted as $d(i,j)$ in the discussion of
Sec.~\ref{sec_Ising}.
In Fig.~\ref{fig:DM} we report $\r_d$ 
at different points in the $(J/U,\mu/U)$ plane, crossing the
insulator-superfluid transition around the tip of the first lobe.

In the MI phase $\r_d$ decays to zero, while in the SF phase we checked that
it correctly decays to $|\la a \ra|^2$. The decay is always quite fast, and in
the SF phase we do not have enough points to perform a reliable fit to extract
a correlation length, altought it is clear that the decay becomes slower and slower
on approaching
the transition to the MI. On the contrary, in the MI phase it is very easy
to extract the correlation length by measuring the slope of $\log \r_d$ versus
$d$ for $d \geq 3$. The result is plotted in the inset of Fig.~\ref{fig:DM}
and shows that the correlation length increases also on the MI side.

It might be surprising at first sight that the correlation length stays finite
at the transition; however, as we discussed at the end of section~\ref{sec_Ising},
this is a general property of Bethe lattices that is
related to the fact that the number of sites $\NN_d$ 
at distance $d$ from a given one
scales as $(c-1)^d$ for large $d$. 
Despite this difference at criticality, 
we believe that the possibility of investigating spatial
correlations on the Bethe lattice is very interesting 
away from critical points, and is an important improvement of the cavity
method with respect to the mean field approximation.

\section{Detailed derivation}
\label{sec_derivation}
\subsection{Lattice boson models in the Suzuki-Trotter formalism}

This section is devoted to a complete derivation of the equations used in
Sec.~\ref{sec_qcav_number}. We shall consider a slightly more general
Hamiltonian than (\ref{eq_def_bh}),
\beq
H = - \sum_\ij J_\ij ( a^\dag_i a_j + a^\dag_j a_i ) 
+ \sum_{i=1}^N V_i(a^\dag_i a_i ) \ ,
\label{eq_def_bh_generalized}
\eeq
where the hopping strength $J_\ij$ is allowed to vary from edge to edge and
we consider arbitrary local potential energies $V_i(n)$. This will be
convenient both for notational reasons and because part of the following
discussion will also apply to the disordered Bose-Hubbard model.
The original model
(\ref{eq_def_bh}) is recovered by taking $J_\ij = J$ on all links, and 
$V_i(n) = \frac{U}{2}n(n-1) - \mu n$. The Hilbert space of the model is 
spanned by the occupation number vectors $|n_1,\dots,n_N \rangle$, where 
$n_i$ is a positive or null integer counting the number of particles on 
site $i$. We shall use the more compact notation 
$|\un\rangle = |n_1,\dots,n_N \rangle$ to denote one of these basis vectors.
For completeness we recall the action of the annihilation $a_i$ and creation
$a^\dag_i$ operators in this basis, 
\bea
a_i |\un \rangle &=& \sqrt{n_i} 
| n_1,\dots,n_{i-1},n_i-1,n_{i+1},\dots, n_N\rangle \ , \nonumber \\
a^\dag_i |\un \rangle &=& \sqrt{n_i+1} 
| n_1,\dots,n_{i-1},n_i+1,n_{i+1},\dots, n_N\rangle \ .
\label{eq_aadag_onn}
\eea
The number operators $a^\dag_i a_i$ are diagonal in this basis, with 
$a^\dag_i a_i | \un \rangle = n_i | \un \rangle$.
The partition function of the Hamiltonian (\ref{eq_def_bh_generalized})
will be computed using the Suzuki-Trotter formula (\ref{eq_Suzuki}),
cutting the imaginary time axis of length $\beta$ in a number $\Ns$ 
(ultimately sent to infinity) of slices. At each of these slices a 
representation of the identity in the occupation number basis is inserted.
This yields
\beq
Z = \Tr \left[ e^{-\beta H} \right] = \lim_{\Ns \to\infty} 
\sum_{\un^1, \cdots, \un^\Ns} \exp\left[-\frac{\beta}{\Ns} 
\sum_{\alpha = 1}^\Ns \sum_{i=1}^N V_i (n^\alpha_i)\right]
\prod_{\alpha=1}^\Ns \langle \un^\alpha | 
e^{\frac{\beta}{\Ns} \underset{\ij}{\sum} J_\ij
( a^\dag_i a_j + a^\dag_j a_i ) } | \un^{\alpha+1} \rangle \ .
\label{eq_Z1}
\eeq
The index $\alpha$ is the discrete coordinate between $1$ and $\Ns$ along the
imaginary time axis, and we use periodic boundary conditions 
$\un^{\Ns+1} =\un^1$. The quantum problem has thus been transformed in a
classical one, in terms of the imaginary-time dependent
classical variables $\un^\alpha$. The expression above of the hopping 
interaction is however unpractical for our future needs, we shall thus 
transform it by introducing a set of auxiliary variables 
$\uy^\alpha= \{ y_{i \to j}^\alpha \}$ which
can take values $0$ or $1$, depend on the discrete time $\alpha$, and are 
defined on the directed edges of the graph (an edge $\ij$ thus
bears, for each time $\alpha$, two variables $y_{i \to j}^\alpha$ 
and $y_{j\to i}^\alpha$). It is simple to show that
\begin{multline}
\langle \un^\alpha | 
e^{\frac{\beta}{\Ns} \underset{\ij}{\sum} J_\ij
( a^\dag_i a_j + a^\dag_j a_i ) } | \un^{\alpha+1} \rangle = 
\sum_{\uy^\alpha} \prod_\ij
\left(\frac{\beta J_\ij \sqrt{n_j^{\alpha+1} n_i^\alpha}}
{\Ns}\right)^{y_{i\to j}^\alpha}
\left(\frac{\beta J_\ij \sqrt{n_i^{\alpha+1} n_j^\alpha}}
{\Ns}\right)^{y_{j\to i}^\alpha} \\
\prod_{i=1}^N \I\left(n_i^{\alpha+1} = n_i^\alpha + 
\sum_{j \in \di} [y_{j \to i}^\alpha - y_{i \to j}^\alpha ] \right) 
+ O\left(\frac{1}{\Ns^2} \right) \ ,
\label{eq_trick}
\end{multline}
where here and in the following $\I(A)=1$ if the condition $A$ is fulfilled,
0 otherwise. We take by convention $x^{y=0}=1$ for any value of $x$ (including
zero). The justification of (\ref{eq_trick}) can be done by inspecting the 
behaviour of its left and right-hand-side order by order in $1/\Ns$. The
leading term corresponds to $\un^\alpha = \un^{\alpha +1}$, and indeed all
$y$'s must vanish at this order. 
At order $1/\Ns$ a single $y$, say $y_{i \to j}^\alpha$, is equal 
to 1, meaning that a boson has hopped from site $i$ to site $j$ between the 
discrete times $\alpha$ and $\alpha+1$. The term 
$\sqrt{n_j^{\alpha+1} n_i^\alpha}$ follows from the action of 
creation/annihilation operators recalled in (\ref{eq_aadag_onn}). The kinetic
energy part of the partition function (\ref{eq_Z1}) can be transformed by
using the representation (\ref{eq_trick}) for each time $\alpha$; reordering
the various terms leads to a compact expression,
\beq
Z = \lim_{\Ns \to\infty} \sum_{\ubn,\uby} \prod_\ij w_\ij (\by_\ij)
\prod_{i=1}^N w_i(\bn_i,\{ \by_\ij \}_{j \in \di}) \ .
\label{eq_Z2}
\eeq
Following the convention we already used in Sec.~\ref{sec_overview} bold
symbols stand for time dependent quantities, with a discrete time coordinate
$\alpha \in [1,\Ns]$ as long as $\Ns$ is finite, or a continuous time 
$\t \in [0,\b]$ in the limit $\Ns \to \infty$. The correspondence
between the two notations is $\t = \a \b/\Ns$. We also use the
shorthand $y_\ij^\alpha=(y_{i \to j}^\alpha,y_{j \to i}^\alpha)$ for the pair
of hopping variables on both directions of an edge $\ij$. The weight of a
configuration $(\ubn,\uby)$ factorizes into a product over all
edges of 
\beq
w_\ij (\by_\ij) = \prod_{\alpha=1}^\Ns
\left(\frac{\beta J_\ij}{\Ns}\right)^{y^\alpha_{i\to j} + y^\alpha_{j \to i}}
\ ,
\label{eq_w_ij}
\eeq
the hopping strength $J_\ij$ is thus conjugated to the number of hopping 
events between sites $i$ and $j$. The second part of the weight in 
Eq.~(\ref{eq_Z2}) is a product over the sites of
\begin{multline}
w_i(\bn_i,\{\by_\ij \}_{j \in \di}) =
\exp\left[-\frac{\beta}{\Ns} \sum_{\alpha = 1}^\Ns V_i(n^\alpha_i)\right] \\
\prod_{\alpha=1}^{\Ns}\left\{ \I\left(n_i^{\alpha+1} = n_i^\alpha + 
\sum_{j \in \di} [y_{j \to i}^\alpha - y_{i \to j}^\alpha ] \right) 
\left(\sqrt{n_i^{\alpha+1}}\right)^{\underset{j \in \di}{\sum} 
y^\alpha_{j \to i}}
\left(\sqrt{n_i^\alpha}\right)^{\underset{j \in \di}{\sum} 
y^\alpha_{i \to j}} \right\} \ .
\label{eq_w_i}
\end{multline}
The first line arises directly from the local potential energy term of the
Hamiltonian. The second line enforces the consistency of the occupation number 
trajectory $n_i^\alpha$ with the hopping events that occur on the adjacent 
edges, and incorporates the $\sqrt{n}$ terms arising from the action of the
creation/annihilation operators.

The expression of the partition function in terms of a classical model with 
imaginary time-dependent variables we just obtained is valid for any 
underlying graph. We want to emphasize that the spatial structure of the 
classical model is the same as the original one: the variables $\bn_i$ are
located on the vertices of the graph, the $\by_\ij$ on the edges, and in the
weights $w_i$ the interactions only occur between edges and site variables
which were adjacent in the original model. In particular, if the latter
was locally tree-like, this remains true for the representation (\ref{eq_Z2})
on which the Bethe approximation can be performed, as we shall do in the next
subsection.

\subsection{The Bethe free energy}

One can follow two roads to obtain the self-consistent equation 
(\ref{eq_fpoint_number}). The first one corresponds to the reasoning
presented in the simpler case of the Ising model in Sec.~\ref{sec_Ising}.
Assuming the graph to be a tree, one can easily compute recursively
$Z_{i \to j}(\by_\ij)$, the partial partition function for the subtree rooted
at $i$ excluding $j$, for a given value of the hopping trajectory $\by_\ij$.
Introducing its normalized counterpart 
$\eta_{i \to j}(\by_\ij) = Z_{i \to j}(\by_\ij)/z_{i \to j}$, one obtains
\beq
\eta_{i \to j}(\by_\ij) = \frac{1}{z_{i \to j}}
w_\ij (\by_\ij)
\sum_{\bn_i,\{\by_\ik \}_{k\in\dimj} }
w_i(\bn_i,\by_\ij,\{ \by_\ik \}_{k \in \dimj})
\prod_{k \in \dimj} \eta_{k \to i}(\by_\ik) \ .
\label{eq_bh_msg}
\eeq
The probability distribution of the occupation number trajectory on site $i$
is then expressed as
\beq
\eta_i(\bn_i) = \frac{1}{z_i} \sum_{\{\by_\ij \}_{j \in \di}}
w_i(\bn_i,\{ \by_\ij \}_{j \in \di})
\prod_{j \in \di} \eta_{j \to i}(\by_\ij) \ .
\label{eq_final_sg}
\eeq
These last two equations are the analogs of 
Eq.~(\ref{eq_msg_Ising}) of the Ising
model. If one considers now the homogeneous case of a Bethe lattice of
connectivity $c$, with the same hopping strength $J_\ij=J$ on all edges and
$V_i(n) = V(n)$ for all sites, the probability laws $\eta_{i \to j}$ take
a common value $\eta_{\rm cav}$ on all directed edges, which is the solution
of Eq.~(\ref{eq_fpoint_number}). In that equation $\by$ is an hopping 
trajectory on an edge, characterized more explicitly as
$\by=(y_+^1,y_-^1,\dots,y_+^\Ns,y_-^\Ns)$, where $y_+^\alpha=1$ 
(resp. $y_-^\alpha=1$) means a particle has arrived (resp. leaved) the 
considered site at discrete time $\alpha$. The expression of the weights 
$w_{\rm link}$ and $w_{\rm iter}$ of Eq.~(\ref{eq_fpoint_number}) is found by
specializing (\ref{eq_w_ij}) and (\ref{eq_w_i}) to the homogeneous situation,
\beq
w_{\rm link}(\by) = \left(\frac{\beta J}{\Ns} 
\right)^{\underset{\alpha=1}{\overset{\Ns}{\sum}} [y_+^\alpha+y_-^\alpha] } \ ,
\label{eq_wlink}
\eeq
and
\begin{multline}
w_{\rm iter}(\by,\by_1\dots,\by_{c-1}) = \sum_\bn e^{-\frac{\beta}{\Ns} 
\overset{\Ns}{\underset{\alpha=1}{\sum}} V(n^\alpha)} \\
\prod_{\alpha=1}^{\Ns}\left\{ \I\left(n^{\alpha+1} = n^\alpha + 
y_-^\alpha-y_+^\alpha +\sum_{i=1}^{c-1} [y_{+,i}^\alpha - y_{-,i}^\alpha ] \right) 
\left(\sqrt{n^{\alpha+1}}\right)^{y_-^\alpha + \overset{c-1}{\underset{i=1}{\sum}} 
 y_{+,i}^\alpha}
\left(\sqrt{n^\alpha}\right)^{y_+^\alpha + \overset{c-1}{\underset{i=1}{\sum}} 
y_{-,i}^\alpha} \right\} \ .
\label{eq_witer}
\end{multline}
Note that the direction of $\by_i$, for $i\in[1,c-1]$, is reversed with 
respect to $\by$, hence the inversion of the signs on the variation of 
$n^\alpha$ from one discrete time to the next. Finally the weight
$w_{\rm site}$ used in Eq.~(\ref{eq_number_final}) to obtain the probability
distribution of an occupation number trajectory reads
\beq
w_{\rm site}(\bn,\by_1\dots,\by_c) = e^{-\frac{\beta}{\Ns} 
\overset{\Ns}{\underset{\alpha=1}{\sum}} V(n^\alpha)} 
\prod_{\alpha=1}^{\Ns}\left\{ \I\left(n^{\alpha+1} = n^\alpha + 
\sum_{i=1}^c [y_{+,i}^\alpha - y_{-,i}^\alpha ] \right) 
\left(\sqrt{n^{\alpha+1}}\right)^{\overset{c}{\underset{i=1}{\sum}} 
 y_{+,i}^\alpha}
\left(\sqrt{n^\alpha}\right)^{\overset{c}{\underset{i=1}{\sum}} 
y_{-,i}^\alpha} \right\} \ .
\label{eq_wsite}
\eeq

The second strategy to reach the equation (\ref{eq_fpoint_number}) on
$\eta_{\rm cav}$ consists in first writing the Bethe 
approximation~\cite{Yedidia} for
the free-energy of the model defined in (\ref{eq_Z2}) for an arbitrary
graph. This yields
\bea
F = -\frac{1}{\beta}\ln Z = &-& \frac{1}{\beta}\sum_{i=1}^N \ln\left( 
\sum_{\bn_i,\{ \by_\ij \}_{j \in \di}} 
w_i(\bn_i,\{ \by_\ij \}_{j \in \di}) \prod_{j \in \di}
\eta_{j \to i}(\by_\ij )
\right) \nonumber \\ &+& \frac{1}{\beta}
\sum_\ij \ln \left( \sum_{\by_\ij} \frac{1}{w_\ij(\by_\ij)}
\eta_{i \to j}(\by_\ij) \eta_{j \to i}(\by_\ij)  \right) \ .
\label{fVAR_sg}
\eea
One can check that this relation is exact whenever the graph is a tree,
otherwise it corresponds to the Bethe approximation. An important property
of this expression of the free-energy is its variational character: the
stationarity conditions with respect to the parameters $\eta_{i \to j}$ are
nothing but the local recurrence equations (\ref{eq_bh_msg}). This means
that the computation of the derivatives of the free-energy with respect
to its parameters $\mu,\beta,J_\ij,\dots$ can be performed by considering
only the explicit dependence on these parameters, discarding the implicit
dependence through the $\eta_{i \to j}$. We shall also show in 
Appendix~\ref{sec_static} how to devise a further (static) approximation
based on this variational expression of the free-energy.
Considering the particular case of
an homogeneous Bethe lattice leads to the following free-energy per site:
\beq\label{fVAR}
f = \lim_{N \to \infty }\frac{-1}{\beta N} \ln Z = -\frac{1}{\beta}
\ln\left(\sum_{\bn,\by_1,\dots,\by_c}  
w_{\rm site}(\bn,\by_1,\dots,\by_c) 
\eta_{\rm cav}(\by_1)\dots\eta_{\rm cav}(\by_c)\right) 
+ \frac{c}{2\beta} \ln\left(
\sum_{\by} \frac{1}{w_{\rm link}(\by)} \eta_{\rm cav}^2(\by) \right) \ ,
\eeq
where we have used the fact that a regular graph of $N$ vertices and 
connectivity $c$ has $cN/2$ edges. One can see that (\ref{eq_fpoint_number})
is the stationarity condition of (\ref{fVAR}) with respect to $\eta_{\rm cav}$.

By using the recurrence equations (\ref{eq_bh_msg}) it is possible to
rewrite the Bethe approximation of the free-energy, for an arbitrary graph,
as
\beq
F = -\frac{1}{\beta}\ln Z = -\frac{1}{\beta} 
\sum_\ij \ln \left(\sqrt{z_{i \to j} z_{j \to i}} \right)
+ \frac{1}{\beta} \sum_{i=1}^N \frac{c_i-2}{2} \ln z_i \ ,
\eeq
where $c_i$ is the degree of vertex $i$, and the various $z$'s correspond to
the normalizations in (\ref{eq_bh_msg}),(\ref{eq_final_sg}). This form does not
have the variational property explained above; it has however the advantage
of being easier to compute numerically. In particular for the homogeneous
case it yields
\beq
f= \lim_{N \to \infty }\frac{-1}{\beta N} \ln Z = 
-\frac{c}{2\beta} \ln z_{\rm cav}
+ \frac{c-2}{2\beta} \ln z_{\rm site} \ , 
\label{eq_f_nonvar}
\eeq
where $z_{\rm cav}$ and $z_{\rm site}$ are the normalizations in, respectively,
Eqs.~(\ref{eq_fpoint_number}) and (\ref{eq_number_final}).

\subsection{The resolution of the cavity equation}

As we have explained in Sec.~\ref{sec_qcav_number_eq} we use a weighted sample
representation of $\eta_{\rm cav}$ to solve (\ref{eq_fpoint_number}). The
point we want to specify more precisely here is the method used to sample
hopping trajectories from the conditional law $P(\by|\by_1,\dots,\by_{c-1})$ 
and to compute the associated normalization $\Z(\by_1,\dots,\by_{c-1})$ (both
defined in Eq.~(\ref{eq_PZ})). A similar construction can be found
for the slightly simpler case of quantum spin 1/2 models in~\cite{qcav_our}.

Let us begin with the computation of $\Z$. 
A first important remark, already mentioned
above, is that in the continuous time limit ($\Ns\to \infty$) the hopping
trajectories $\by$ typically contain only a finite number (with respect to
$\Ns$) of hopping events, i.e. of discrete times 
$\alpha$ where $y_\pm^\alpha \neq 0$. We can thus assume without loss of generality
that the hopping events for the different trajectories occur at different
values of the discrete time. Let us call $p$ the total number of hopping events
occuring in $(\by_1,\dots,\by_{c-1})$, and denote $\alpha_1<\dots<\alpha_p$
their discrete time of occurence. We consider the Hilbert space of a single
site, with $a$ and $a^\dag$ the annihilation/creation operators, and call
$b_j=a$ (resp. $b_j=a^\dag$) if the hopping of time $\alpha_j$
is towards (resp. outside) the vertex under consideration. Finally we define
$c_\alpha = b_j$ when $\alpha=\alpha_j$, $c_\alpha = \I$ (the identity
operator) otherwise. A moment of thought reveals that
\beq
\Z(\by_1,\dots,\by_{c-1}) = 
\sum_{\by} w_{\rm link}(\by) w_{\rm iter}(\by,\by_1,\dots,\by_{c-1})= 
\sum_\bn \prod_{\alpha=1}^\Ns \langle n^\alpha |
e^{\frac{\beta}{\Ns}(-V(a^\dag a)+ J(a+a^\dag) )} c_\alpha 
| n^{\alpha+1} \rangle \ ,
\eeq
up to corrections of order $\Ns^{-2}$. The sum over $\bn$ can thus be written
as a trace of this product of operators. To perform the continuous time limit
it is convenient to define $\t_j = \frac{\beta}{\Ns}\alpha_j$, which are
the continuous times of the hopping events in $(\by_1,\dots,\by_{c-1})$.
We also introduce $\hW(\lambda) = e^{\lambda(-V(a^\dag a)+ J(a+a^\dag) )}$, 
the propagator of an imaginary time evolution on an interval of length 
$\lambda$ for a single site Hamiltonian $V(a^\dag a) - J(a+a^\dag)$. We thus
obtain finally
\beq
\Z(\by_1,\dots,\by_{c-1}) = 
\Tr \left( \hW(\t_1) b_1 \hW(\t_2-\t_1) b_2 \dots\hW(\t_p-\t_{p-1}) b_p \hW(\beta-\t_p) \right) 
\ .
\label{eq_Z}
\eeq
The single site Hilbert space is a priori infinite dimensional because
the number of particles is not bounded. Very high occupation numbers are
however unlikely because of their prohibitive 
potential energy. We can thus safely
put a cutoff $N_{\rm cut}$ on the possible values of $n$, which will not
change the properties of the model if $N_{\rm cut}$ is sufficiently large with
respect to the average density of particles. This would amount formally to
take $V(n)=+\infty$ for $n \ge N_{\rm cut}$. In this case the dimension of the
Hilbert space is reduced to $N_{\rm cut}$, and (\ref{eq_Z}) is nothing but the
trace of the product of 
a finite number of finite size matrices, whose numerical evaluation
does not present any particular difficulty.

We turn now to the problem of the generation of an hopping trajectory $\by$
given the ones of the $c-1$ other neighbors, according to the law
$P(\by|\by_1,\dots,\by_{c-1})$ stated in (\ref{eq_PZ}). As explained in
Sec.~\ref{sec_qcav_number_eq} (see in particular Fig.~\ref{fig_traj_iter}),
we can determine $\by$ by first drawing the occupation number trajectory $\bn$
and then deduce $\by$ from its discontinuities not associated to hopping
events in $(\by_1,\dots,\by_{c-1})$. We stick to the notation
$\t_1 < \dots <\t_p$ and $b_1,\dots,b_p$ for the parametrization of the
hopping events in $(\by_1,\dots,\by_{c-1})$. Let us call $n_0=n(\t=0)$,
$n_i$ (resp. $n'_i$)
the value of $n(\t)$ at a time just after $\t_i$ (resp. just before
$\t_{i+1}$), with the conventions $n'_p=n_0$. The joint probability law
of these occupation numbers which arise from the expressions of $w_{\rm link}$
and $w_{\rm iter}$ given in Eqs.~(\ref{eq_wlink}), (\ref{eq_witer}) reads in the
continuous time limit
\beq
P(n_0,n'_0,n_1,n'_1,\dots,n_p|\by_1,\dots,\by_{c-1}) = 
\frac{1}{\Z(\by_1,\dots,\by_{c-1})}
\langle n_0 | \hW(\t_1) | n'_0 \rangle  
\prod_{i=1}^p \left\{ \langle n'_{i-1}|b_i|n_i \rangle
\langle n_i | \hW(\t_{i+1}-\t_i) | n'_i \rangle \right\} \ ,
\eeq
with $\t_{p+1}=\beta$. This probability law is well normalized according to
the above expression of $\Z(\by_1,\dots,\by_{c-1})$. Moreover because of the
``unidimensional'' structure of the imaginary time axis it is rather simple to
generate a set of integer numbers $(n_0,n'_0,n_1,n'_1,\dots,n_p)$ according to
this law. Once these intermediate occupation numbers are known, the generation 
of the occupation trajectory $n(\t)$ can be done independently on each of the
$p+1$ time intervals between the imposed hopping events. On the $i+1$'th
interval between $\t_i$ and $\t_{i+1}$ one has to generate the trajectory
corresponding to a path integral representation of the effective Hamiltonian
$V(a^\dag a) - J(a+a^\dag)$, conditioned to begin in $n_i$ at time $\t_i$ and
to end in $n'_i$ at $\t_{i+1}$.

To simplify the notation let us consider this problem for an interval of
imaginary
time $[0,\lambda]$, with boundary conditions $n(0)=n$, $n(\lambda)=n'$. One
way to justify the sampling procedure is to start with the identity
\beq
e^{\lambda(X+Y)} = e^{\lambda X} + \int_{0}^\lambda \dd \t \ e^{\t X} Y 
e^{(\lambda-\t)(X+Y)} \ ,
\eeq
valid for any non-commutative operators $X$ and $Y$. We apply it with
$X=-V(a^\dag a)$ and $Y=J(a+a^\dag)$, which yields
\bea
\langle n | \hW(\lambda) | n' \rangle &=& \langle n | 
e^{-\lambda V(a^\dag a) } + \int_0^\lambda \dd \t \ e^{-\t V(a^\dag a)} 
J (a+a^\dag) \hW(\lambda -\t) | n' \rangle \\
&=& \delta_{n,n'} e^{-\lambda V(n)} + J \int_0^\lambda \dd \t \ e^{-\t V(n)} 
\langle n | (a+a^\dag) \hW(\lambda -\t) | n' \rangle \ .
\label{eq_matelm}
\eea
The interpretation of this formula is as follows: a path representative of the
matrix element $\langle n | \hW(\lambda) | n' \rangle$ is either a constant
trajectory (possible only if $n=n'$), or it is constant up to a time $\t$,
where it jumps to $n \pm 1$, and is followed by a path of length $\lambda -
\tau$ with boundary conditions $n \pm 1$ and $n'$. We can thus sample the
path $n(\t)$ according to the following recursive procedure:
\begin{itemize}
\item if $n=n'$ and with probability 
$e^{-\lambda V(n)}/\langle n | \hW(\lambda) | n \rangle$, set $n(\t)=n$ for 
$\t\in[0,\beta]$ and exit the procedure
\item otherwise 
\begin{itemize}
\item draw a time $\t \in [0,\lambda]$ with the probability law written
  below in Eq.~(\ref{eq_G})
\item draw $\sigma= 1$ with probability 
$\frac{\langle n | a \hW(\lambda -\t) | n' \rangle}
{\langle n | (a+a^\dag) \hW(\lambda -\t) | n' \rangle}$, $\sigma=-1$ otherwise
\item set $n(\t') = n$ for $\t' \in [0,\t]$
\item call the same procedure with initial condition $n+\sigma$, final 
condition $n'$, for the imaginary time interval $[\t,\lambda]$
\end{itemize}
\end{itemize}
The probability law for the time $\t$ of the first discontinuity in the number
occupation trajectory follows from (\ref{eq_matelm}); it is more convenient to
express it as the probability $G(u)$ that $\t\le u$,
\beq
G(u) = \frac{\int_0^u \dd \t \ e^{-\t V(n)} 
\langle n | (a+a^\dag) \hW(\lambda -\t) | n' \rangle}
{\int_0^\lambda \dd \t \ e^{-\t V(n)} 
\langle n |(a+a^\dag) \hW(\lambda -\t) | n' \rangle} \ .
\label{eq_G}
\eeq
Because of the cutoff on the possible number occupation one can numerically
compute this function by diagonalizing the effective Hamiltonian (only once
for each choice of the parameters $U,\mu,J$). Drawing $\t$ then amounts to use
a random number $G$ uniformly in $[0,1]$ and set $\t=u^{-1}(G)$ (the
functional inverse of $G(u)$).

\subsection{The computation of the observables}

We will show now that, as announced in Sec.~\ref{sec_qcav_number_eq}, all the
physical observables can be obtained as averages over $\eta_{\rm cav}$ of
some functions of the hopping trajectories. The computation of such an average
can then be done as a simple sampling of the population
representing $\eta_{\rm cav}$, as explained in Eq.~(\ref{eq_avg_g}).

Let us begin with the free-energy. Using the formula (\ref{eq_f_nonvar}) its
computation amounts to the determination of the normalization constants
$z_{\rm cav}$ and $z_{\rm site}$. The former reads 
\beq
z_{\rm cav} = \sum_{\by_1,\dots,\by_{c-1}} \eta_{\rm cav}(\by_1) \dots 
\eta_{\rm cav}(\by_{c-1}) \Z(\by_1,\dots,\by_{c-1}) \ ,
\eeq
which is readily evaluated since we obtained an explicit form of
$\Z(\by_1,\dots,\by_{c-1})$ in Eq.~(\ref{eq_Z}). The latter is seen from
(\ref{eq_number_final}) to be
\beq
z_{\rm site} = \sum_{\by_1,\dots,\by_c} \eta_{\rm cav}(\by_1) \dots 
\eta_{\rm cav}(\by_c) \sum_{\bn} w_{\rm site}(\bn,\by_1,\dots,\by_c) \ .
\eeq
Using again the notation
$\t_1 < \dots <\t_p$ and $b_1,\dots,b_p$ for the parametrization of the
hopping events in $(\by_1,\dots,\by_c)$ and
exploiting the expression (\ref{eq_wsite}) for $w_{\rm site}$, we obtain
in the continuous-time limit 
\beq
z_{\rm site} = \sum_{\by_1,\dots,\by_c} \eta_{\rm cav}(\by_1) \dots 
\eta_{\rm cav}(\by_c) \Tr \left( 
\hW_0(\t_1) b_1 \hW_0(\t_2-\t_1) b_2 \dots b_p \hW_0(\beta-\t_p) \right) ,
\eeq
where we defined $\hW_0(\lambda) = e^{-\lambda V(a^\dag a)}$, the single-site
propagator without its hopping term. Indeed all the possible hopping events to
and from the considered site are fixed by the $c$ trajectories
$\by_1,\dots,\by_c$. This trace is over a product of finite size matrices
(thanks to the bound $N_{\rm cut}$ on the occupation numbers) and hence
computationally harmless. Note that only configurations of $\by_1,\dots,\by_c$
which encode the same number of jumps towards/outside the central site do
contribute, because $\hW_0$ is diagonal in the number basis.

We consider now the thermal average $\la q(a^\dag_i a_i) \ra$ of an arbitrary
function $q(n)$ of the number operator on one site. From the expression
(\ref{eq_number_final}) of the probability distribution $\eta(\bn)$ we obtain
\beq\label{eq_av_qaa}
\la q(a^\dag_i a_i ) \ra = \frac{1}{z_{\rm site}} \sum_{\by_1,\dots,\by_c} 
\eta_{\rm cav}(\by_1) \dots \eta_{\rm cav}(\by_c) \Tr \left( q(a^\dag a)
\hW_0(\t_1) b_1 \hW_0(\t_2-\t_1) b_2 \dots b_p \hW_0(\beta-\t_p) \right) \ .
\eeq
This formula allows to compute the density of particles, the average local
energy $e_L$ and the occupation probability $P_{n_0}$ by taking $q(n)=n$, 
$q(n)=V(n)$ and $q(n) = \delta_{n,n_0}$, respectively.

The order parameter $\la a_i \ra$ is obtained by the insertion of an
annihilation operator in the effective single-site problem,
\beq
\la a_i \ra = \frac{1}{z_{\rm site}} \sum_{\by_1,\dots,\by_c} 
\eta_{\rm cav}(\by_1) \dots \eta_{\rm cav}(\by_c) \Tr \left( a
\hW_0(\t_1) b_1 \hW_0(\t_2-\t_1) b_2 \dots b_p \hW_0(\beta-\t_p) \right) \ .
\label{eq_res_a}
\eeq
One has indeed by definition $\la a_i \ra = \Tr[a_i e^{-\beta H}]/Z$, where the
trace is here over the Hilbert space of the $N$-sites Hamiltonian. One can
reproduce all the steps leading to the representation (\ref{eq_Z2}) of the
partition function, with the additional operator $a_i$ modifying the
expression of the weight $w_i$ on the considered site. The other sites are
however left unmodified, and hence integrating over them leads to the same
equation on $\eta_{\rm cav}$. The presence of $a_i$ only shows up when the sum
over the degrees of freedom of site $i$ is performed, and leads to this
insertion of the annihilation operator in the single-site computation
(\ref{eq_res_a}). We now come back on the problem of the initial condition
in the population dynamics mentioned in Sec.~\ref{sec_qcav_number_eq}. One can
indeed see that (\ref{eq_res_a}) strictly vanishes whenever all hopping
trajectories in the support of $\eta_{\rm cav}$ have the same number of jumps
in the two directions of their edge. Moreover this symmetry is conserved by
the iterative equation (\ref{eq_fpoint_number}); we thus had to initialize the
population dynamics algorithm including asymmetric hopping trajectories. 
In the ``insulating'' phase these trajectories disappear during the iterations,
while in the ``superfluid'' a finite fraction of them keeps the asymmetry,
thus allowing for a non zero value of $\la a_i \ra$.

The computation of the Green function 
$G^i_>(\t) = \Tr [ e^{-(\b-\t) H} a_i e^{-\t H} a^\dag_i]/Z$
can be done similarly, with now the insertion of two creation/annihilation 
operators in the single-site problem. 
Similarly to Eq.~(\ref{eq_av_qaa}), and using the same notations, 
one obtains
\beq\begin{split}
G_>(\t) =& \frac{1}{z_{\rm site}} \sum_{\by_1,\dots,\by_c} 
\eta_{\rm cav}(\by_1) \dots \eta_{\rm cav}(\by_c) \times \\
&\times  \Tr \left( a
\hW_0(\t_1) b_1 \hW_0(\t_2-\t_1) b_2 \dots  b_i \hW_0(\t-\t_i) a^\dag \hW_0(\t_{i+1}-\t) b_{i+1} \cdots  b_p \hW_0(\beta-\t_p) \right) \ ,
\end{split}\eeq
where the index $i$ is such that $\t_i < \t < \t_{i+1}$.

We turn now to the computation of the average kinetic energy 
$e_K = \la H_K \ra /N$. By definition it is equal to the derivative of the
free-energy per site with respect to $J$. Using the variational property of
the expression (\ref{fVAR}), we can thus write
\beq
e_K = - \frac{c}{2 \beta J} \frac{\underset{\by}{\sum} 
\frac{1}{w_{\rm link}(\by)} \eta_{\rm cav}^2(\by) |\by| }
{\underset{\by}{\sum} \frac{1}{w_{\rm link}(\by)} \eta_{\rm cav}^2(\by) } \ ,
\eeq
where we have defined $|\by|$ as the total number of hopping events in $\by$.
Indeed the only explicit dependence on $J$ in (\ref{fVAR}) is in the weight
$w_{\rm link}$ given in Eq.~(\ref{eq_wlink}). Replacing one of the 
$\eta_{\rm cav}$ by its expression (\ref{eq_fpoint_number}) both in the
numerator and the denominator, and reordering terms leads to
\beq
e_K = - \frac{c}{2 \beta J} \frac{1}{z_{\rm site}} \sum_{\by_1,\dots,\by_c} 
\eta_{\rm cav}(\by_1) \dots \eta_{\rm cav}(\by_c) |\by_1| \Tr \left(
\hW_0(\t_1) b_1 \hW_0(\t_2-\t_1) b_2 \dots b_p \hW_0(\beta-\t_p) \right) \ .
\eeq

\begin{figure}
\begin{center}
\includegraphics[width=10cm]{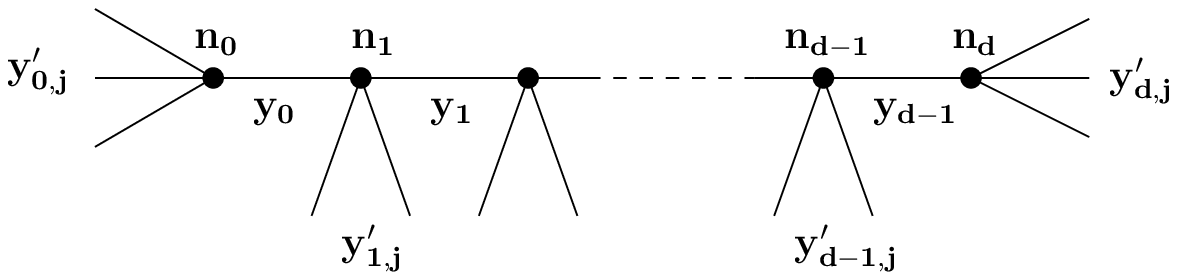}
\end{center}
\caption{Computation of the one-particle density matrix at distance $d$,
$\rho_d = \la a^\dag_0 a_d \ra$.}
\label{fig_chain}
\end{figure}

Finally we consider the computation of the one-particle density matrix 
$\rho_d$ for two sites at distance $d$. For notational simplicity we
shall call them $0$ and $d$, $\rho_d=\la a^\dag_0 a_d \ra$. 
The unique path linking these two sites is
schematized in Fig.~\ref{fig_chain}; we have distinguished the hopping
trajectories $\by_i$ along the chain for $i \in [0,d-1]$, from the
contribution of the edges outside the chain, $\by'_{i,j}$. The index $j$ runs
from 1 to $c_i$, with $c_0=c_{d}=c-1$ and $c_1=\dots=c_{d-1}=c-2$, as the 
extremities $0$ and $d$ have one less neighbor in the chain. The
computation of $\la a^\dag_0 a_d \ra$ is a generalization of the determination
of $\la a_i \ra$ explained above: one writes 
$\la a^\dag_0 a_d \ra = \Tr[a^\dag_0 a_d e^{-\beta H}]/Z$ 
and expresses the numerator
with the Suzuki-Trotter formula where the local weights $w_0$ (resp. $w_d$ are
modified with the insertion of a creation (resp. annihilation) operator. 
Summing over all degrees of freedom outside the chain of sites between $0$ and
$d$ leads to
\beq
\la a^\dag_0 a_d \ra = \frac{
\underset{\{\by'_{i,j}\}}{\sum}
\underset{i,j}{\prod} \eta_{\rm cav}(\by'_{i,j})
\underset{\{\by_i \}}{\sum} 
\underset{i}{\prod} w_{\rm link}(\by_i)
\underset{\{\bn_i\}}{\sum} w_{\rm site}^+(\bn_0,\by_0,\{\by'_{0,j}\})
\underset{i}{\prod} w_{\rm site}(\bn_i,\by_{i-1},\by_i,\{\by'_{i,j}\})
w_{\rm site}^-(\bn_d,\by_{d-1},\{\by'_{d,j}\})
}{
\underset{\{\by'_{i,j}\}}{\sum}
\underset{i,j}{\prod} \eta_{\rm cav}(\by'_{i,j})
\underset{\{\by_i \}}{\sum} 
\underset{i}{\prod} w_{\rm link}(\by_i)
\underset{\{\bn_i\}}{\sum} w_{\rm site}(\bn_0,\by_0,\{\by'_{0,j}\})
\underset{i}{\prod} w_{\rm site}(\bn_i,\by_{i-1},\by_i,\{\by'_{i,j}\})
w_{\rm site}(\bn_d,\by_{d-1},\{\by'_{d,j}\})
} \ ,
\label{eq_densmat}
\eeq
where in the numerator $w^\pm$ are modifications of the vertex 
weight~(\ref{eq_wsite}) due to the insertion of the creation/annihilation 
operators on sites $0$ and $d$. Integrating progressively the degrees of
freedom along the chain one can easily shows that the denominator of 
(\ref{eq_densmat}) equals $z_{\rm cav}^{d} z_{\rm site}$. To compute the
numerator we shall define a sequence of probability distributions
$\eta_{\rm cav}^{(i)}$, as the law of the hopping trajectories $\by_i$ along
a chain where the part between $i+1$ and $d$ is removed. This can be computed
by recurrence on $i$, with
\beq
\eta_{\rm cav}^{(0)}(\by) = \frac{1}{z_{\rm cav}^{(0)}} w_{\rm link}(\by)
\sum_{\by_1,\dots,\by_{c-1}} 
\eta_{\rm cav}(\by_1) \dots \eta_{\rm cav}(\by_{c-1}) 
\sum_\bn w_{\rm site}^+(\bn,\by,\by_1,\dots,\by_{c-1}) \ , 
\eeq
and
\beq
\eta_{\rm cav}^{(i+1)}(\by) = \frac{1}{z_{\rm cav}^{(i+1)}} w_{\rm link}(\by)
\sum_{\by_1,\dots,\by_{c-1}} \eta_{\rm cav}^{(i)}(\by_1)
\eta_{\rm cav}(\by_2) \dots \eta_{\rm cav}(\by_{c-1}) 
\sum_\bn
w_{\rm site}(\bn,\by,\by_1,\dots,\by_{c-1}) \ . 
\eeq
These probability distributions can be encoded numerically as weighted samples,
in the same way as we did for $\eta_{\rm cav}$. Once they have been determined
up to distance $d-1$ we can compute
\beq
z_{\rm site}^{(d)} = \sum_{\by_1,\dots,\by_c} \eta_{\rm cav}^{(d-1)}(\by_1)
\eta_{\rm cav}(\by_2) \dots \eta_{\rm cav}(\by_c) 
\sum_\bn w_{\rm site}^-(\bn,\by_1,\dots,\by_c) \ .
\eeq
One then finds
\beq
\la a^\dag_0 a_d \ra = \frac{z_{\rm site}^{(d)}}{z_{\rm site}}
\prod_{i=0}^{d-1} \frac{z_{\rm cav}^{(i)}}{z_{\rm cav}} \ .
\eeq

\section{Conclusions}
\label{sec_conclu}

In this paper we have extended the analytical tools of the quantum cavity method
to Bethe lattice bosonic models.
The cavity approach yields a self-consistent equation for the probability
of hopping trajectories in continuous imaginary time, which can be
efficiently solved numerically using the population dynamics algorithm.
Though this numerical step is unavoidable to obtain quantitative predictions
it is distinct from a Quantum Monte Carlo algorithm, in particular the
thermodynamic limit is taken analytically.

We have shown how this method allows to compute easily 
several physical observables such as thermodynamic quantities 
(average occupation number,
energy, free energy, condensate fraction, \ldots)
as well as imaginary time Green functions. Let us emphasize in particular
that computing the free-energy (which is often difficult with
other methods, requiring for instance thermodynamic integration in Monte Carlo
simulations) is crucial to determine the location of first-order
phase transitions which can occur in generalized versions of the model.

The finite connectivity of the
Bethe lattice leads to two kind of improvements with respect to the mean-field
description: a distance between two sites can be defined,
which allows the computation of spatial correlation functions. Moreover
the description of the Mott Insulator phase is richer, with a non-trivial
effect of the hopping of the particles. The prediction for the
zero-temperature phase diagram of Bethe lattices with connectivity
4 and 6 is in reasonable agreement with the Quantum Monte Carlo simulations
in 2 and 3 dimensions~\cite{bh_qmc_2d,bh_qmc_3d}.

Let us sketch some possible directions for future work. A first possibility
would be to explore the next levels of the Cluster Variation Method~\cite{CVM}
hierarchy of approximations, which contains the mean-field and the Bethe
approximations as its first two steps. This approach, similar in spirit to
the Cluster DMFT~\cite{CDMFT}, should improve the accuracy of the
predictions for small dimensions. It should also be interesting to investigate
the effect of disorder in the Bethe lattice model; if the Bose Glass 
phase~\cite{FiFi,inguscio1} is destroyed at the mean-field level, it could be possible
to describe it on the Bethe lattice, which can indeed display localization
properties~\cite{localization_Bethe}. Another open direction is the study
of several generalizations of the Bose-Hubbard model, including bosons with
spin, multiple interacting species of 
particles~\cite{BDMFT_1,BDMFT_2,bh_multi,inguscio2}, and interactions between 
nearest-neighbor sites~\cite{bh_extended}. In particular 
in the latter case, we expect that the presence of interactions inducing 
geometrical frustration
will give rise to glass phases, since the same happens in the classical
limit of zero hopping~\cite{BM}. It would be very interesting to check
whether these glass phases could exhibit Bose-Einstein Condensation as
it happens for their crystalline counterparts~\cite{supersolid_lattice}. This would add some
new element to the ongoing discussion on the relevance of disorder to
induce superfluidity~\cite{supersolid_disorder}, and might be
important to interpret and understand recent experiments 
on supersolidity~\cite{supersolid_exp}.

We finally point out that, although in this paper we have only focused on
the very low temperature regime (except for the transition temperature at 
unit filling displayed in Fig.~\ref{fig_PD_Tpos}), the method presented 
here can be applied at higher temperature (and is even numerically easier 
in that case), which can be useful for the interpretation of cold atoms 
experiments~\cite{bh_finite_T}. From this 
perspective the case of unidimensional systems should deserve a special
attention, as they fall in the category of tree structure and could thus
be solved with the cavity method, even in the presence of a site-dependent 
chemical potential modelizing the trapping field.

\acknowledgments

The method we used in this paper has been first developed for quantum spins
in collaboration with F.~Krzakala and A.~Rosso~\cite{qcav_our} whom we
warmly thank. We also acknowledge useful comments on a preliminary version of the
manuscript from G.~Biroli, K.~Byczuk, G.~Carleo, N.~Dupuis, J.~Est\`eve, F.~Gerbier, D.~Vollhardt 
and M.~Schir\`o, and M.~M\'ezard and R.~Thomale for interesting discussions.

\appendix

\section{Static approximation}
\label{sec_static}

We have obtained in (\ref{fVAR_sg}) a variational expression of the Bethe
free-energy, in terms of the parameters $\h_{i \to j}(\by_\ij)$. A further
approximation consists in restricting the $\h_{i \to j}$ to a simply 
parametrized subspace, and finding the extremum of (\ref{fVAR_sg}) within
this simpler ansatz. A possible choice for this parametrization, inspired
by the so-called static approximation developed for quantum spin 
models~\cite{static,qcav_our}, consists in retaining in $\by_\ij$ only the 
information on the number and direction of the jumps, but assuming their times
of occurence to be uniformly distributed. This corresponds in formula to the
ansatz
\beq
\h_{i \to j}(\by_\ij) = \sum_{n_{i \to j},n_{j \to i}=0}^\io 
\r_{i \to j}(n_{i \to j},n_{j \to i}) 
\frac{1}{\binom{\Ns}{n_{i \to j}} \binom{\Ns}{n_{j \to i}} } 
\I \left( \sum_\a y_{i \to j}^\a = n_{i \to j} \right) 
\I \left( \sum_\a y_{j \to i}^\a = n_{j \to i} \right) \ ,
\eeq
where $\r_{i \to j}$ is the distribution probability for the number of jumps
from $i$ to $j$ and viceversa. The two terms of the variational 
free-energy (\ref{fVAR_sg}) can be computed within this ansatz.
One obtains indeed in the limit $\Ns \to \io$:
\beq\label{s1}
\sum_{\by_\ij} \frac{1}{w_\ij(\by_\ij)}
\eta_{i \to j}(\by_\ij) \eta_{j \to i}(\by_\ij) = 
\sum_{n_{i \to j},n_{j \to i}=0}^\io 
\r_{i \to j}(n_{i \to j},n_{j \to i})\r_{j \to i}(n_{j \to i},n_{i \to j})
\frac{n_{i \to j}! n_{j \to i}!}{(\b J_\ij)^{n_{i \to j} + n_{j \to i}}}
\eeq
and
\beq\label{s2}
\sum_{\bn_i,\{ \by_\ij \}_{j \in \di}} 
w_i(\bn_i,\{ \by_\ij \}_{j \in \di}) \prod_{j \in \di}
\eta_{j \to i}(\by_\ij )
=  \sum_{ \{n_{i\to j},n_{j \to i} \}_{j \in \di}} 
\prod_{j \in \di} \r_{j \to i}(n_{j \to i},n_{i \to j})
A_i\left( \sum_{j \in \di} n_{j \to i} ,\sum_{j \in \di} n_{i \to j}\right) \ ,
\eeq
where
\beq
 A_i(N_+,N_-) =
\frac{1}{\b^{N_+ + N_-}} \left. 
\frac{d^{N_+}}{d z_+^{N_+}} \frac{d^{N_-}}{d z_-^{N_-}} 
\Tr  \left[ e^{-\b ( V_i(a^\dag a) - z_+ a^\dag - z_- a ) } \right] 
\right|_{z_+ = z_- = 0} \ .
\eeq
As a simple illustration of this static approximation one can recover the
mean-field free-energy (\ref{eq_f_mf}) for the case of a Bethe lattice
of connectivity $c$ with uniform hopping strength $J$ and the same potential
energy $V(n)$ on all sites. This is indeed obtained after a short computation
by taking a Poisson distribution of average $\beta J \lambda$ for the 
number of jumps in each direction,
\beq
\r(n_+,n_-) = e^{-2\b J \l} \frac{(\b J \l)^{n_+ + n_-}}{n_+! n_-!} \ ,
\eeq
where $\l$ is proportional to the parameter $\psi$ in (\ref{eq_f_mf}).


\end{document}